%% file: 0arxiv.tex
\documentclass[sigconf]{acmart}
\usepackage{multirow}

\renewcommand\footnotetextcopyrightpermission[1]{}
\setcopyright{none}

\input{00macros}

\includeversion{arxiv}
\excludeversion{submission}

\begin{document}

\title{\method: Verifiable ML Property Cards using Hardware-assisted Attestations}

\author{Vasisht Duddu}
\affiliation{%
  \institution{University of Waterloo}
  \country{Canada}
}
\email{vasisht.duddu@uwaterloo.ca}

\author{Oskari Järvinen}
\affiliation{%
  \institution{Aalto University}
  \country{Finland}
}
\email{oskari.jarvinen@aalto.fi}

\author{Lachlan J. Gunn}
\affiliation{%
  \institution{Aalto University}
  \country{Finland}
}
\email{lachlan@gunn.ee}

\author{N. Asokan}
\affiliation{%
  \institution{University of Waterloo}
   \country{Canada}
}
\email{asokan@acm.org}

\begin{abstract}
\input{01abstract}
\end{abstract}

\maketitle
\pagestyle{plain}

\setlist{nolistsep}
\input{1introduction}
\input{2background}
\input{3problem}

\input{4attestations} 
\input{5framework}

\input{6evaluation}
\input{7related}
\input{8discussions}

\begin{arxiv}
\input{00ack}
\end{arxiv}

\bibliographystyle{ACM-Reference-Format}
\bibliography{paperS}

\end{document}

%% file: 00macros.tex
\usepackage{tikz}
\usepackage{dblfloatfix}
\usepackage[inline]{enumitem}
\usepackage{xspace}
\usepackage{amsmath}
\usepackage{booktabs}
\usepackage[hyphenbreaks]{breakurl}
\usepackage{multirow}
\usepackage{amsfonts}
\usepackage{soul}

\usepackage{subfigure}
\usepackage[table]{colortbl}
\usepackage{versions}
\usepackage{bbm}
\usepackage{pifont}
\usepackage{tikz}

\usepackage{cleveref}
\usepackage[ruled,vlined,noresetcount]{algorithm2e}
\usepackage{caption}

\usetikzlibrary{shapes,arrows}
\usetikzlibrary{backgrounds, positioning, fit}
\makeatletter
\tikzset{
    database/.style={
        path picture={
            \draw (0, 1.5*\database@segmentheight) circle [x radius=\database@radius,y radius=\database@aspectratio*\database@radius];
            \draw (-\database@radius, 0.5*\database@segmentheight) arc [start angle=180,end angle=360,x radius=\database@radius, y radius=\database@aspectratio*\database@radius];
            \draw (-\database@radius,-0.5*\database@segmentheight) arc [start angle=180,end angle=360,x radius=\database@radius, y radius=\database@aspectratio*\database@radius];
            \draw (-\database@radius,1.5*\database@segmentheight) -- ++(0,-3*\database@segmentheight) arc [start angle=180,end angle=360,x radius=\database@radius, y radius=\database@aspectratio*\database@radius] -- ++(0,3*\database@segmentheight);
        },
        minimum width=2*\database@radius + \pgflinewidth,
        minimum height=3*\database@segmentheight + 2*\database@aspectratio*\database@radius + \pgflinewidth,
    },
    database segment height/.store in=\database@segmentheight,
    database radius/.store in=\database@radius,
    database aspect ratio/.store in=\database@aspectratio,
    database segment height=0.1cm,
    database radius=0.25cm,
    database aspect ratio=0.35,
}
\makeatother

\renewcommand*{\algorithmcfname}{Attestation}

\Crefname{algocf}{Attestation}{Attestations}

\newenvironment{infattIO}[1][htb]
  {\renewcommand{\algorithmcfname}{Attestation}
   \begin{algorithm}[#1]%
  }{\end{algorithm}}

\newenvironment{infattRob}[1][htb]
  {\renewcommand{\algorithmcfname}{Attestation}
   \begin{algorithm}[#1]%
  }{\end{algorithm}}

\newenvironment{infattFair}[1][htb]
  {\renewcommand{\algorithmcfname}{Attestation}
   \begin{algorithm}[#1]%
  }{\end{algorithm}}

\newenvironment{infattAcc}[1][htb]
  {\renewcommand{\algorithmcfname}{Attestation}
   \begin{algorithm}[#1]%
  }{\end{algorithm}}

\newenvironment{trattDist}[1][htb]
  {\renewcommand{\algorithmcfname}{Attestation}
   \begin{algorithm}[#1]%
  }{\end{algorithm}}

\newenvironment{trattPoT}[1][htb]
  {\renewcommand{\algorithmcfname}{Attestation}
   \begin{algorithm}[#1]%
  }{\end{algorithm}}

\newcommand{\method}{\textsc{Laminator}\xspace}

\newcommand{\verifier}{$\mathcal{V}$\xspace}
\newcommand{\prover}{$\mathcal{P}$\xspace}
\newcommand{\modelarch}{\mathcal{M}_{ar}\xspace}
\newcommand{\trainconfig}{\mathcal{T}\xspace}
\newcommand{\model}{\mathcal{M}\xspace}
\newcommand{\modelprov}{\mathcal{M}\xspace}
\newcommand{\dtestprov}{\mathcal{D}^{prov}_{te}\xspace}
\newcommand{\dtrainprov}{\mathcal{D}^{prov}_{tr}\xspace}
\newcommand{\dtrain}{\mathcal{D}_{tr}\xspace}
\newcommand{\dtest}{\mathcal{D}_{te}\xspace}
\newcommand{\drob}{\mathcal{D}_{rob}\xspace}
\newcommand{\accuracy}{\mathcal{A}_{acc}\xspace}
\newcommand{\accuracyRob}{\mathcal{A}_{rob}\xspace}
\newcommand{\accuracyFair}{\mathcal{A}_{fair}\xspace}
\newcommand{\accuracyVer}{\mathcal{A}^{ver}_{acc}\xspace}
\newcommand{\accuracyRobVer}{\mathcal{A}^{ver}_{rob}\xspace}
\newcommand{\accuracyFairVer}{\mathcal{A}^{ver}_{fair}\xspace}
\newcommand{\out}{\mathcal{O}\xspace}
\newcommand{\inp}{\mathcal{I}\xspace}
\newcommand{\property}{\textsc{p}\xspace}

%% file: 01abstract.tex
Regulations increasingly call for various assurances from machine learning (ML) model providers about their training data, training process, and model behavior.
For better transparency, industry (e.g., Huggingface and Google) has adopted \emph{model cards} and \emph{datasheets} to describe various properties of training datasets and models.
In the same vein, we introduce the notion of \emph{inference
cards} to describe the properties of a given inference (e.g., binding of the output to the model and its corresponding input).
We coin the term \emph{ML property cards} to collectively refer to these various types of cards.

To prevent a malicious model provider from including false information in ML property cards, they need to be \textit{verifiable}. 
We show how to construct \emph{verifiable ML property cards} using \emph{property attestation}, technical mechanisms by which a prover (e.g., a model provider) can attest to various ML properties to a verifier (e.g., an auditor).
Since prior attestation mechanisms based purely on cryptography are often narrowly focused (lacking versatility) and inefficient, we need an efficient mechanism to attest different types of properties across the entire ML model pipeline.

Emerging widespread support for \emph{confidential computing} has made it possible to run and even train models inside hardware-assisted \emph{trusted execution environments} (TEEs), which provide highly efficient attestation mechanisms.
We propose \method, which uses TEEs to provide the \textit{first} framework for verifiable ML property cards via \emph{hardware-assisted ML property attestations}.
\method is \emph{efficient} in terms of overhead, \emph{scalable} to large numbers of verifiers, and \emph{versatile} with respect to the properties it can prove during training or inference\begin{arxiv}\footnote{ACM Conference on Data and Application Security and Privacy (CODASPY), 2025.}\end{arxiv}.

%% file: 1introduction.tex
\section{Introduction}\label{sec:introduction}


Machine learning (ML) models are increasingly being deployed for various high-stakes applications like medical diagnosis, bank loans, and autonomous vehicles~\cite{christian2021alignment}. This has raised concerns about the various risks to data privacy, fairness and robustness~\cite{trustAISurvey}. 
Further, several emerging regulations strive to ensure that the training process and model behavior during inference are as expected~\cite{ec2019ethics,nist,eureg,congress,eur,dpia,ico,whitehouse}. 
For instance, a model provider may be required to convince a verifier (e.g., a regulator or potential customer) that a model meets the expected level of accuracy, fairness, privacy or robustness~\cite{ec2019ethics,whitehouse,ico,dpia,eur} or  distributional properties of training data comply with regulations~\cite{eureg,congress}. Privacy or confidentiality considerations may imply that model providers need to do so \emph{without disclosing the model or the training data to the verifier}.
The ability to demonstrate regulatory compliance \emph{without compromising confidentiality} of models or training data will enable the development of a marketplace for ML models.

Several companies, including Google and Huggingface, have adopted \emph{model cards}~\cite{mitchell2019model} and \emph{datasheets}~\cite{gebru2021datasheets,pushkarna2022data} for better transparency of ML model and dataset properties, respectively. We propose a similar notion called \emph{inference cards} for describing the properties of a specific \emph{inference}, e.g., binding the output of an inference, to the input and the model. 
Collectively, we refer to them as \emph{ML property cards} which are short documents containing information about \emph{ML properties} such as intended use, risks and limitations, training procedure, datasets, and metrics for evaluation (e.g., accuracy, fairness, robustness)~\cite{mitchell2019model}.
These ML property cards are generated by the model provider who trains and/or deploys the model, implicitly assuming that model providers are trusted to include correct information in ML property cards.  But this may not be a valid assumption, as it has been demonstrated that malicious entities can distribute models via marketplaces~\cite{mithrilsecurityPoisonGPTPoison}.
 Hence, \emph{ML property cards} must be \emph{verifiable} to ensure that \emph{their} information is \emph{neither false nor tampered with}.


Huggingface offers a utility that allows model providers to send their model to a \emph{trusted certifier} (e.g., Huggingface) to compute some metrics. The trusted certifier then creates a property card with a signature to verify that the metric is valid.
However, this is not easily-extensible, being limited only to metrics that the certifier can compute. Hence, it is not possible to prove
\begin{enumerate*}[label=\roman*),itemjoin={,\xspace}]
\item properties of sensitive datasets that cannot be disclosed to the certifier (e.g., demographic distribution, or certifying provenance of training dataset)
\item that a model was trained using a specific optimization (e.g., differential privacy~\cite{abadi2016deep}, or group fairness~\cite{hardt}), or
\item that an output is the result of the model applied on a corresponding input during inference.
\end{enumerate*}
We propose using \emph{ML property attestation}, first introduced in our previous work~\cite{duddu2023attesting}, which are technical mechanisms by which a prover (e.g., a model provider) can demonstrate relevant properties (training-time and inference-time) to a verifier (e.g., a client querying a model deployed as a service, a customer purchasing a model, or a regulatory authority).

Current ML property attestation schemes have various drawbacks in terms of accuracy~\cite{duddu2023attesting}, efficiency~\cite{duddu2023attesting,zkpTraining}, or scalability to multiple verifiers~\cite{duddu2023attesting}.
Further, prior works are limited to specific types of attestation such as proof of training (that a model was trained on some dataset)~\cite{zkpTraining,zkpTraining2,shamsabadi2023confidentialprofitt,zkDL}, distributional property attestations (proving distributions of attributes in training data)~\cite{duddu2023attesting,holmes}, or showing that a model was trained with a specific optimization~\cite{confidentialdpproof,fairaudit1,fairaudit2,fairaudit3,yadav2024fairproof}.
No existing property attestation mechanism is simultaneously \textbf{efficient}, \textbf{scalable}, and \textbf{versatile}.

Trusted Execution Environments (TEEs) already have the notion of hardware-assisted \emph{remote attestation} to prove local system or software configuration to a remote verifier. The trusted computing research community has extended these mechanisms to more general property attestation~\cite{propatt}.
TEEs are widely available in commercially-available \emph{confidential computing} environments~\cite{confidential-computing}. Recent developments by several hardware vendors~\cite{amx,h100}, have made it possible to run and train ML models efficiently inside TEEs. We take advantage of the increasing use of TEEs in ML deployments to realize efficient, scalable and versatile ML property attestation.

We propose the first software framework for \emph{hardware-assisted ML property attestation}. 
We use TEEs to establish \textit{bindings} between key properties of training dataset and the model during inference and training. These bindings are signed by the TEE's secret attestation key, generating ML property attestations. When combined with signed external certificates from trusted entities, these attestations enable verifiers to draw conclusions about models, their behavior, or their training datasets. These attestations can be then be used for verifiable ML property cards. We claim the following contributions: we
\begin{enumerate}[leftmargin=*]
\item propose \emph{inference cards} which include properties specific to each inference (e.g., binding input-model-output) and the collective of \emph{ML property cards} to include model cards, datasheets, and inference cards (Section~\ref{sec:attestations})
\item propose \method\footnote{Code: \url{https://github.com/ssg-research/laminator}.}, the \textit{first} framework to generate \emph{verifiable ML property cards} using hardware-assisted attestations throughout training and inference (Section~\ref{sec:framework})
\item using four different types of attestations, we demonstrate that \method is efficient, scalable, and versatile. (Section~\ref{sec:evaluation})
\end{enumerate}

%% file: 2background.tex
\section{Background}\label{sec:background}

We present the ML notations used in this work (Section~\ref{sec:mlback}), details about model and dataset cards (Section~\ref{sec:propcards}), and trusted execution environments (Section~\ref{sec:teeback}).

\subsection{Machine Learning Notations}\label{sec:mlback}

Consider an ML model configuration file $\modelarch$ describing the model architecture including the number and types of layers, and a training configuration file $\trainconfig$ which includes the objective function, learning rate, weight decay, and other hyperparameters such as number of epochs. 
Additionally, consider a dataset $\dtrain\sim\mathbb{D}_{\property}$ where $\mathbb{D}_{\property}$ is an underlying distribution satisfying some distributional properties $\property$ (for example, the ratio of females in the distribution). 
Since, we use i.i.d sampling to get $\dtrain$, it satisfies $\property$. 
We obtain model $\model$ by training as per the description in $\trainconfig$ using $\modelarch$ on $\dtrain$, i.e., $\model\leftarrow \text{Train}(\dtrain, \modelarch, \trainconfig)$. 
During inference, $\model$ takes an input $\inp$ and output $\out$, i.e., $\out$ $\leftarrow$ $\model$($\inp$). Additionally, $\model$ is evaluated with respect to accuracy, fairness or robustness on some unseen test dataset $\dtest$ sampled from same distribution as $\dtrain$. 
We assume that both $\dtrain$ and $\dtest$ are composed of data records $\inp$, the corresponding classification label $y$, and some sensitive attribute $z$.
We summarize notations used in the rest of the paper in Table~\ref{tab:notations}.

\begin{table}[htb]
    \centering
    \caption{Summary of notations and their descriptions.}
    \footnotesize
    \begin{tabular}{l|p{4.5cm}}
    \hline
    \textbf{Notation}  & \textbf{Description} \\
    \hline
    \prover & Prover\\
    \verifier & Verifier\\
    $\property$ & Distributional property\\
    $\modelarch$ & Model architecture\\
    $\trainconfig$ & Training Configuration\\
    $\inp$ & Input to a model \\
    $y$ & Classification label\\
    $z$ & Sensitive attribute \\
    $\out$ & Output from a model for $\inp$ \\
    $\modelprov$ & Prover's model to be attested \\
    $\dtrainprov$ & Prover's training dataset\\
    $\dtestprov$ & Prover's test dataset\\
    $\drob$ & Test dataset with adversarial examples \\
    $\accuracy$ & Test accuracy\\
    $\accuracyRob$ & Robust accuracy\\
    $\accuracyFair$ & Fairness metric\\
    $\accuracyVer$ & Verifier's measured accuracy\\
    $\accuracyRobVer$ & Verifier's measured robust accuracy\\
    $\accuracyFairVer$ & Verifier's measured fairness metric\\
    $h(\cdot)$ & cryptographic hash function\\
    \hline
    \end{tabular}
    \label{tab:notations}
\end{table}

\subsection{Model Cards and Datasheets}\label{sec:propcards}

To improve transparency into ML training and inference, companies have adopted \emph{model cards} and \emph{datasheets}. We describe them below.

\noindent\textbf{Model cards} are short documents containing additional information about different static model properties~\cite{mitchell2019model}. 
These include properties prior to deployment of the model as a service to clients to query via an API reference.
They were designed to provide better transparency into the design of ML model which includes training and evaluation of the model prior to deployment for inference.
These have been extensively adopted by different companies such as Google~\cite{modelcardsGoogleCloud} and Huggingface~\cite{huggingfaceModelCards}. Model cards contain a variety of additional information about ML properties: 
\begin{enumerate*}[label=\roman*),itemjoin={,\xspace}]
\item \emph{general properties} including information about the model's intended use, risks such as bias, and other limitations
\item \emph{training properties} including training datasets used, training hyperparameters, and model architecture
\item \emph{evaluation properties} including information about the different test datasets and metrics used for measuring accuracy, precision, recall, fairness, and robustness.
\end{enumerate*}

\noindent\textbf{Datasheets} describe different properties corresponding to the training dataset ~\cite{gebru2021datasheets,pushkarna2022data}. Datasheets, also known as \emph{dataset cards}, have been used extensively by companies such as Huggingface~\cite{huggingfaceDatasetCards}. They describe collection process and sources from which the raw data was collected, pre-processing including cleaning and labelling to obtain training dataset from raw data, intended use, composition of dataset, and demographic subpopulations and their distribution.

Model cards and datasheets are created by model providers who train the model and/or deploy them as a service. Hence, we have to trust the model providers for the information claimed in the cards.

\subsection{Trusted Execution Environments (TEEs)}\label{sec:teeback}

TEEs~\cite{tee} provide an environment in which security-critical code (trusted applications, or TAs) can run without outside interference from other TAs, or the rest of the system (the Rich Execution Environment, or REE, which contains the operating system and normal software).
TEEs can be described in terms of three main properties: isolation, secure storage, and remote attestation~\cite{tee}. Isolation refers to the inability for outside software to interfere with the operation of a TA, or to read its confidential state. 
Secure storage refers to persistent storage made available to TAs that resists interference by other software. 
Remote attestation refers to the ability of a TA to prove some aspect of a TEE's state to a remote party, e.g., that the secret key of a particular keypair is held by a TA whose code has a particular cryptographic hash.

This work uses Intel's Software Guard Extensions (SGX)~\cite{sgx} in its evaluation, though any TEE technology that offers isolation and remote attestation can equally be used. SGX provides enclave-based TEEs, allowing applications to create a TEE (an \emph{enclave}) in their own memory space and run a TA of their choice. SGX ensures that neither the host application, the OS, nor external code can access the enclave's memory, with communication occurring through defined entry/exit points and via memory read/write operations with the host application.
SGX enclaves can remotely attest themselves by obtaining a \emph{quote}, which serves as a signature over TA-provided data and certifies the enclave's identity. This identity includes \texttt{MRENCLAVE} (a hash of the TA's initial memory), \texttt{MRSIGNER} (a hash of the public key that signed the TA), user data (a chunk of data provided by the TA), debug mode status, versioning, and other data~\cite[EREPORT]{sgx}.


A verifier can assess the validity of a quote, then use the hash in its \texttt{MRENCLAVE} and knowledge of the behavior of the corresponding code to determine the TA's state when the quote was generated. If a TA produces quotes only for specific computations, the verifier can confirm that the computation was performed and the quoted result was produced. This enables proving properties of data and computations within the TA.

\subsection{Gramine Library OS}\label{sec:gramine}

Gramine\footnote{https://gramineproject.io/} allows to run unmodified applications inside an SGX enclave using a library OS, enabling the use of off-the-shelf software such as Python. 
This requires a manifest file which describes the application's security configuration, including information on files that will be accessible, with or without integrity checking.
By default, only files explicitly listed as \emph{trusted files} (whose data is incorporated into the enclave hash at build time, and have their integrity checked on access), or \emph{allowed files} (which can be accessed but are not subject to integrity checks) are accessible from within the enclave.
This happens when the Gramine tooling automatically hashes the trusted files and adds them to the enclave manifest, which is then incorporated into the SGX enclave's~\texttt{MRENCLAVE}.
Thus, marking files as trusted prevents them from being replaced by a malicious host, as the hash verification will fail. This enables even security-critical files to be stored outside the enclave.

%% file: 3problem.tex
\section{Problem Statement}\label{sec:problem}

We introduce the term \emph{ML property cards} to refer to transparency declarations like model cards and datasheets used by model providers. Our goal is to design a framework for generating verifiable ML property cards using ML property attestations for training and inference. This ensures the information in the cards is (a) correctly measured and (b) not falsified or tampered with. We begin by presenting the adversary model (Section~\ref{sec:threatmodel}), desiderata for an ideal attestation mechanism (Section~\ref{sec:desiderata}), and limitations of current attestation mechanisms (Section~\ref{sec:limitations}).

\subsection{Adversary Model}\label{sec:threatmodel}

We consider a prover \prover that trains, evaluates, or deploys a model $\modelprov$.
We assume \prover has their own train and test dataset indicated as $\dtrainprov$ and $\dtestprov$.
During training, \prover wants to attest different properties of $\modelprov$'s training to a verifier \verifier. For instance, attesting that $\modelprov$ was trained using $\trainconfig$, $\modelarch$, and $\dtrain$ (a.k.a. ``proof of training'').
After training, \prover evaluates $\modelprov$ to check that it satisfies some minimal accuracy, fairness and robustness requirements prior to deployment. 
Once \prover deploys $\modelprov$ as a service, clients can query $\modelprov$ by sending input $\inp$ via an API interface and obtain the corresponding output $\out$. Here, \prover attests that $\out$ was generated by $\modelprov$ for a specific $\inp$. 
In Section~\ref{sec:infcards}, we expand the notion of ML property cards to include a new category, \emph{inference cards} to cover this type of inference-time attestation.

\verifier may not trust \prover. If \prover can subvert the attestation mechanism, it can fool \verifier into believing that some property holds when in fact it does not. For instance, \prover can make false claims despite not using the expected $\trainconfig$ or $\modelarch$, using a different dataset than $\dtrain$ which does not satisfy $p$, incorrect values for accuracy, fairness, and robustness, or using a different $\out$ for inference. 
We assume that \prover's entire computing platform except for the TEE is untrusted, including the privileged code such as the operating system and hypervisor, which are controlled by \prover.

We assume two roots of trust, signature verification keys belonging to (a) the manufacturer of the TEE (e.g., Intel) who certifies the keys used by enclaves to sign attestations, and (b) trusted certifiers (e.g., CIFAR) that provide attestations of computational properties (with trust in the property grounded in the trusted certifier rather than the TEE), and certificates of non-computational properties, e.g., data provenance.
As long as \verifier trusts them, \verifier can trust any attestation or certificates linked to them.

\subsection{Desiderata for ML Property Attestations}\label{sec:desiderata}

We identify the following desiderata for ML property attestation:
\begin{enumerate}[label=\textbf{R\arabic*},leftmargin=*,leftmargin=*,wide,  labelindent=0pt]
\item \label{efficient}\textbf{Efficient} (low-overhead generation of attestations).
\item \label{scalable}\textbf{Scalable} (supports large numbers of provers/verifiers).
\item \label{versatile}\textbf{Versatile} (supports a wide variety of properties; requires minimal effort to add support for new properties).
\end{enumerate}
Our goal is to design a framework to generate verifiable ML property cards that can satisfy all the above requirements.

We assume an attestation mechanism that allows us to execute a piece of code, transforming an input into an output while guaranteeing a certain assertion. 
This guarantee is evidenced by a signature $\operatorname{Sign}_\textrm{att}(\texttt{'Assertion'}, \cdots)$ generated by the attestation mechanism only if the assertion holds. Here, $\operatorname{Sign}_\textrm{att}(\cdot)$ represents a signature created using a private key that is not available outside the mechanism.

In the rest of the paper, we will describe attestations in the form shown in \Cref{att:attestation-example}.
{\renewcommand{\algorithmcfname}{Attestation}
\begin{algorithm}[h!]
\caption{Sample attestation}\label{att:attestation-example}
\textbf{Input}: $i_1, i_2, \ldots$ \\
\textbf{Computation}: $o_1, o_2, \ldots \overset{\$}{\leftarrow} f(i_1, i_2, \ldots)$ \\
\textbf{Output}: $o_1, o_2, \ldots, \operatorname{Sign}_\textrm{att}(\texttt{'AttX'}, g(i_1, i_2, \ldots, o_1, o_2, \ldots))$ \\
\textbf{Assertion}: $P(i_1, i_2, \ldots, o_1, o_2, \ldots)$ \\
\emph{where} $a \overset{\$}{\leftarrow} \operatorname{Proc}()$ denotes probabilistic assignment (i.e.~assigning to $a$ the value of a possibly-probabilistic procedure $\operatorname{Proc}()$).  Not all attestations necessarily include all fields.
\end{algorithm}
}
\Cref{att:attestation-example} corresponds to the assertion that inputs $i_1, i_2, \ldots$ are transformed into outputs $o_1, o_2, \ldots$ by a (possibly probabilistic) procedure $f$, yielding output $\operatorname{Sign}_\textrm{att}(\texttt{'AttX'}, \cdot)$, so long as the assertion $P(i_1, i_2, \ldots, o_1, o_2, \ldots)$ holds.  These attestations will be used as the basis for \method.

\subsection{Limitations of Prior Work}\label{sec:limitations}

ML property cards can be made verifiable using cryptographic techniques like zero-knowledge proofs (ZKPs) or secure multiparty computation (MPC) for verifiable training~\cite{zkDL,shamsabadi2023confidentialprofitt,zkpTraining,zkpTraining2,eisenhofer2022verifiable} and inference~\cite{zkCNN,zenCompiler,Weng2021MystiqueEC,zkMLaaS,veriTrain,safetyNets}.
But no prior scheme meets all three requirements.
\begin{enumerate}[leftmargin=*,leftmargin=*,wide,  labelindent=0pt]
\item \textbf{ZKPs} are \emph{scalable} and can support multiple verifiers. However, they are computationally expensive and do not efficiently extend to neural networks.
For instance, proof of training using ZKPs for a simple logistic regression model on a dataset of $2^{18}$ data records with 1024 attributes, and a batch size of 2014, takes 72 seconds \textit{for one epoch} including training and proof generation time~\cite{zkpTraining}.
Finally, ZKPs lack versatility, as they only support properties that can be adapted to the ZKP scheme~\cite{zkpTraining}, limiting their use for complex functions (e.g., nonlinear activation).
\item \textbf{MPC} requires online interactions between the prover and verifier, meaning that it lacks scalability and efficiency, and requires retraining the model for each verification~\cite{duddu2023attesting}.
\end{enumerate}
Despite ongoing advances in cryptographic techniques~\cite{SoKCrypto}, purely cryptographic protection of many ML operations (e.g., training), remain inefficient. 
Recent optimizations for TEEs to run ML workloads efficiently~\cite{h100,H100-performance,amx} as well as secure outsourcing protocols to GPUs (e.g.,~\cite{goten,mo2022sok}) offer efficient solutions by combining TEEs with hardware accelerators.
In light of these, we explore the use of TEE remote attestation for ML properties, with negligible overhead compared to a baseline setting where an ML application is already running in a confidential computing setting.
We conjecture that using TEE attestation, instead of mechanisms like ZKP, makes verification more efficient. 
We discuss other TEE-based work which focus on confidentiality instead of integrity in Section~\ref{sec:discussions}.

%% file: 4attestations.tex
\section{Verifiable ML Property Cards}\label{sec:attestations}

Current ML property cards describe properties about datasets in the form of datasheets, and about models in the form of model cards. 
However, they currently do not include properties \emph{during each inference} after deployment. 
To this end, we propose the notion of \emph{inference cards} (Section~\ref{sec:infcards}).
We then present different ML property attestations for different verifiable ML property cards (Section~\ref{sec:datasheet},~\ref{sec:modelcards}, and~\ref{sec:attinfcards}) by carrying out operations inside a TEE and then including an attestation in the ML property card.
Finally, we discuss how \verifier can combine different attestations to make meaningful conclusions (Section~\ref{sec:chain}).

\subsection{Inference cards}\label{sec:infcards}

Inference cards include properties for each inference, for instance, showing that an output was obtained from a specific model with a particular input and, when using an explainable model, that an explanation derives from the same inference~\cite{explSurvey}.
Prior works using non-interactive ZKPs for verifiable inference in ML models~\cite{zkCNN,zkMLaaS,zenCompiler} and large language models~\cite{sun2024zkllm}, aim to provide similar functionality to inference cards.


\input{figures/fig_overview}

\subsection{Attestations for Datasheets}\label{sec:datasheet}

We first present dataset attestations required for datasheets: distributional property attestation of $\dtrain$.
We hereafter refer to a hash function as $h(\cdot)$. 
All the frequently used notations are in Table~\ref{tab:notations}.

\noindent\textbf{Distributional Property Attestation} attests that distributional properties $\property$ (such as gender ratio) of $\dtrain$ by providing an assertion that it has some value which can later be checked with some requirements by \verifier~\cite{duddu2023attesting}.
This can either be done during training~\cite{duddu2023attesting,zkpTraining} or later during inference in combination with \Cref{att:prooftr}~\cite{duddu2023attesting}. We can have different distributional properties across:
\begin{enumerate*}[label=\roman*),itemjoin={,\xspace}]
\item a single attribute (e.g., male-to-female ratio)
\item multiple attributes (e.g., male-to-female \textit{and} caucasian-to-non-caucasian ratios)
\item joint properties (i.e., fraction of data records for a given classification label for different subgroups).
\end{enumerate*}
For a given $\dtrainprov$, \method specifies the properties to be attested and generates the following: $\property$, \texttt{h}($\dtrainprov$), and $sign_{sk}$(\texttt{h}($\dtrainprov$), $\property$)). We present this in \Cref{att:propatt}.

\begin{trattDist}
\caption{Distributional Property Attestation (DistAtt)}\label{att:propatt}
\noindent\textbf{Input}: $\dtrainprov$\\
\noindent\textbf{Output}: $\property$, $\operatorname{Sign}_\textrm{att}(\texttt{`DistAtt'}, \property, \texttt{h}(\dtrainprov))$ \\
\noindent\textbf{Assertion}: $\property(\dtrainprov)$
\end{trattDist}

\subsection{Attestations for Model Cards}\label{sec:modelcards}

We now cover different attestations for model cards, including: 
\begin{enumerate*}[label=\roman*),itemjoin={,\xspace}]
\item proof of training, which includes attesting $\trainconfig$ and $\modelarch$
\item evaluation attestations, which includes accuracy, fairness, and robustness on $\dtestprov$.
\end{enumerate*}
We indicate these in \colorbox{blue!8}{blue} (Figure~\ref{fig:overview}).

\noindent\textbf{Proof of Training} attests that $\modelprov$ was trained on $\dtrainprov$. This requires training $\modelprov$ using a training configuration $\trainconfig$ on $\dtrainprov$. The attestation in shown in \Cref{att:prooftr}.

\begin{trattPoT}
\caption{Proof of Training (PoT)}\label{att:prooftr}
\noindent\textbf{Input}: $\dtrainprov$, $\trainconfig$\\
\noindent\textbf{Output}: $\modelprov$, $\operatorname{Sign}_\textrm{att}(\texttt{`PoT'}$, $\texttt{h}(\modelprov)$, $\texttt{h}(\modelarch)$, $\texttt{h}(\dtrainprov)$, $\texttt{h}(\trainconfig))$ \\
\emph{where} $\modelprov \leftarrow \operatorname{Train}(\dtrainprov, \trainconfig, \modelarch)$ \\
\end{trattPoT}

\noindent\textbf{Accuracy attestation} indicates that $\modelprov$ has a certain accuracy $\accuracy$ as measured using some dataset $\dtestprov$. This ensures
\begin{enumerate*}[label=\roman*),itemjoin={,\xspace}]
\item there are no false claims about $\modelprov$'s $\accuracy$, and
\item $\modelprov$ was evaluated on (a standardized) $\dtestprov$ with no tampering. 
\end{enumerate*}
\method, in addition to $\accuracy$, generates \texttt{h}($\modelprov$), \texttt{h}($\dtestprov$) along with a signature binding them, i.e., $Sign_\textrm{att}$(\texttt{h}($\modelprov$), \texttt{h}($\dtestprov$), $\accuracy$). We use an indicator function $\mathbb{I}_{\{a = b\}}$ that takes the value $1$ if $a = b$, and $0$ otherwise. We summarize this in \Cref{att:accatt}.

\begin{infattAcc}
\caption{Accuracy attestation (AccAtt)}\label{att:accatt}
\noindent\textbf{Input}: $\modelprov$, $\dtestprov$\\
\noindent\textbf{Output}: \texttt{h}($\modelprov$), \texttt{h}($\dtestprov$), $\accuracy$, $\operatorname{Sign}_\textrm{att}(\texttt{`AccAtt'}, \texttt{h}(\modelprov), \texttt{h}(\dtestprov), \accuracy)$  \\
\textbf{Assertion:} $\accuracy = \frac{1}{N}\sum_{(x,y)\in \dtestprov}^{N} \mathbb{I}_{\{\modelprov(x)==y)\}}$
\end{infattAcc}

\noindent\textbf{Fairness Attestation} extends accuracy attestation to evaluate fairness of $\modelprov$ on $\dtestprov$ using different metrics such as accuracy parity, false positive and negative rates across subgroups. We refer to the fairness metric being evaluated as $\accuracyFair$. We attest that $\modelprov$ satisfies $\accuracyFair$ on $\dtestprov$ in \Cref{att:fairatt}. We use demographic parity as an example, but any fairness metric can be substituted.

\begin{infattFair}
\caption{Fairness Attestation (FairAtt)}\label{att:fairatt}
\noindent\textbf{Input}: $\modelprov$, $\dtestprov$\\
\noindent\textbf{Output}: \texttt{h}($\modelprov$), \texttt{h}($\dtestprov$), $\accuracyFair$,$\operatorname{Sign}_\textrm{att}(\texttt{`FairAtt'}, \texttt{h}(\modelprov), \texttt{h}(\dtestprov), \accuracyFair)$\\
\textbf{Assertion:} $\accuracyFair = |P(\modelprov(X)=0 | S=0) - P(\modelprov(X)=0 | S=1)|$ where $(X, Y, S)$ $\in \dtestprov$
\end{infattFair}

\noindent\textbf{Robustness Attestation} attests that $\modelprov$ satisfies some acceptable level of robustness indicated by \emph{robust accuracy} metric ($\accuracyRob$) on a test dataset containing adversarial examples ($\drob$). 
This proceeds in two steps: first, we generate $\drob$ along with its hash (\texttt{h}($\drob$)) inside the enclave. Since $\drob$ is generated locally, its hash serves as the certificate during verification. This certifies that $\drob$ was created by adding adversarial examples, generated with a perturbation of $\epsilon$, to $\dtestprov$.
In the second step, we use accuracy attestation (\Cref{att:accatt}) but using $\drob$ instead of $\dtestprov$, yielding $\accuracyRob$. 

\begin{infattRob}
\caption{Robustness Attestation (RobustAtt)}\label{att:robatt}
\textbf{Generating and Certifying $\drob$:}\\
\noindent\textbf{Input}: $\dtestprov$, $\accuracyRob$, $\epsilon$\\
\noindent\textbf{Output}: $\drob$, \texttt{h}($\drob$), \texttt{h}($\epsilon$), \texttt{h}($\accuracyRob$), $\operatorname{Sign}_\textrm{att}(\texttt{'RobustAttA'}, \texttt{h}(\modelprov), \texttt{h}(\drob), \accuracyRob)$ \\~\\

\textbf{Attesting Robustness}:\\
\noindent\textbf{Input}: $\modelprov$, $\dtestprov$\\
\noindent\textbf{Output}: \texttt{h}($\modelprov$), \texttt{h}($\drob$), $\accuracyRob$, $\operatorname{Sign}_\textrm{att}(\texttt{`RobustAttB'}, \texttt{h}(\modelprov), \texttt{h}(\drob), \accuracyRob)$\\
\textbf{Assertion:} $\accuracyRob = \frac{1}{N}\sum_{(x,y)\in \drob}^{N} \mathbb{I}_{\{\modelprov(x)==y)\}}$
\end{infattRob}

\subsection{Attestations for Inference Cards}\label{sec:attinfcards}

We attest the correctness of input-model-output triples, which can be used to produce inference cards. 

\noindent\textbf{Input-Model-Output Attestation (IOAtt)}, shown in \Cref{att:inpmodout} attests that the output $\out$ was generated from $\modelprov$ for a given input $\inp$. This ensures 
\begin{enumerate*}[label=\roman*),itemjoin={,\xspace}]
\item integrity against malicious \prover who can tamper with $\out$ 
\item non-repudiation to prevent \prover from claiming that $\out$ is not from to $\inp$
\item false claim that $\out'$ is from $\modelprov$ for $\inp$ instead of $\out$. 
\end{enumerate*}

\begin{infattIO}
\caption{Input-Model-Output Attestation (IOAtt)}\label{att:inpmodout}
\textbf{Input:} $\inp$, $\modelprov$\\
\textbf{Output:} $\out$, \texttt{h}($\modelprov$), \texttt{h}($\inp$), \texttt{h}($\out$), $\operatorname{Sign}_\textrm{att}(\texttt{`IOAtt'}, \texttt{h}(\modelprov), \texttt{h}(\inp), \texttt{h}(\out))$ \\
\textbf{Assertion:} $\out = \modelprov(\inp)$
\end{infattIO}



\subsection{Chains of Attestations and Certificates}\label{sec:chain}

Along with the above attestations, certificates from trusted certifiers (e.g., CIFAR) can help verifiers to validate the quality of the training or test datasets. These certificates can certify a \emph{non-computational property} such as e.g.~data provenance or suitability for a particular application.
Combining these certificates with attestations can help \verifier conclude that the model was trained and evaluated on dataset considered good for a particular task.

The assertions of different attestations along with the external certificates by themselves might not be useful in practice. For instance, input-model-output attestation suggests that \textit{for an input $\inp$, $\model$ generates the output $\out$.} However, this information is useful only if \verifier knows how good $\model$ is in terms of accuracy, fairness or robustness.
To use attestations in practice, we chain multiple attestations, used to create verifiable ML property cards, allowing us to arrive at a broader range of conclusions.

As a motivating example, consider a company testing $\modelprov$ for ``good'' accuracy on $\dtest$ using accuracy attestation (\Cref{att:accatt}). This helps \verifier conclude that their $\out$ is indeed from a good model. Including additional attestations, such as fairness (\Cref{att:fairatt}) and robustness (\Cref{att:robatt}), demonstrates that $\modelprov$ meets other application-specific requirements. In all cases, certifying $\dtest$ ensures that metrics are computed on a standardized test dataset certified by a trusted organization.

Combining several training-time attestations (e.g.~\Cref{att:prooftr,att:propatt}) can help to arrive at the following conclusions: \textit{$\model$ was trained on $\dtrain$ satisfying distributional properties $\property$}.
However, we can combine training-time attestation with inference-time attestations to also reach conclusions about the properties of models after training (accuracy, fairness and robustness). Hence, in addition to training attestation, we can additionally prove that \textit{$\out$ was generated from $\model$ for $\inp$, where $\model$ was trained on was trained on $\dtrain$ satisfying $\property$, and $\model$ satisfies the required \{accuracy, fairness, robustness\}}. 

%% file: figures/fig_overview.tex
\pgfdeclarelayer{background}
\pgfdeclarelayer{foreground}
\pgfsetlayers{background,main,foreground}
\tikzstyle{defense} = [rectangle,  minimum width=1.4cm, minimum height=0.5cm, text centered, draw=black]

\begin{figure}[!htb]
\begin{center}
\resizebox{\columnwidth}{!}{ 
\begin{tikzpicture}[line width=1pt]

\node (trdata) [database, fill= cyan!8, align=center, label=below:{\footnotesize Train Dataset ($\dtrain$)}, database radius=0.4cm,database segment height=0.2cm] {};

\node (trmetric) [above of=trdata, fill= orange!8, yshift=1.5cm, align=center] {\footnotesize Data metrics\\\footnotesize (bias, size)};

\draw[->, ultra thick] (trmetric.south) -- (trdata.north);


\node (arch) [defense, fill= blue!8, right of=trdata, xshift=1.4cm, yshift=1cm, align=center, minimum width=1.75cm] {\footnotesize Architecture\\\footnotesize ($\modelarch$)};

\node (trconfig) [defense, fill= blue!8, above of=arch, align=center] {\footnotesize Configuration\\\footnotesize ($\trainconfig$)};

\node (trmodel) [right of=arch, fill= blue!8, draw, xshift=3cm, align=center, minimum width=4cm, minimum height=0.8cm] {\footnotesize Trained Model};


\draw[->, ultra thick] (trconfig.east) -- ([yshift=0.4cm]trmodel.west);
\draw[->, ultra thick] (arch.east) -- (trmodel.west);
\draw[->, ultra thick] (trdata.east) -| ([xshift=-1.7cm]trmodel.south);


\node (tedata) [above of=trmodel, fill= cyan!8, yshift=0.5cm, database,label=above:{\footnotesize Test Dataset ($\dtest$)}, database radius=0.4cm,database segment height=0.2cm] {};
\draw[->, ultra thick] (tedata.south) -- (trmodel.north);


\node (metrics) [align=center, below of=trmodel, yshift=-0.3cm,fill= blue!8] {\footnotesize Accuracy, Fairness,\\\footnotesize Robustness};
\draw[->, ultra thick] (trmodel.south) -- (metrics.north);

\node (inp) [draw, right of=trmodel, fill= red!8, xshift=2cm, yshift=0.5cm, minimum width=1cm] {\footnotesize Input};
\node (out) [draw, fill= red!8, below of=inp] {\footnotesize Output};
\draw[->, ultra thick] (inp.west) -- ([yshift=0.1cm]trmodel.east);
\draw[->, ultra thick] ([yshift=-0.1cm]trmodel.east) -- (out.west);

\node (data) [below of=trdata, yshift=-0.5cm, align=center] {\footnotesize \textbf{Dataset}\\\footnotesize \textbf{Attestation}};
\draw [dashed] (1.35,3) -- (1.35,-2);
\node (tr) [right of=data, xshift=2cm, align=center] {\footnotesize \textbf{Proof of Training}};
\draw [dashed] (5,3) -- (5,-2);
\node (eval) [right of=tr, xshift=2.5cm,align=center, align=center] {\footnotesize \textbf{Evaluation}\\\footnotesize \textbf{Attestation}};
\draw [dashed] (8,3) -- (8,-2);
\node (inf) [right of=eval, xshift=1.5cm,align=center] {\footnotesize \textbf{Inference}\\\footnotesize \textbf{Attestation}};
\end{tikzpicture}
}
\end{center}
\caption{Overview of ML property attestations and their relation to different verifiable model cards: attestations for datasheet (\colorbox{orange!8}{orange}), model card (\colorbox{blue!8}{blue}), inference card (\colorbox{red!8}{red}), along with external certificates (\colorbox{cyan!8}{cyan}).}
\label{fig:overview}
\end{figure}

%% file: 5framework.tex
\section{\method: Framework Design}\label{sec:framework}

We present \method, a framework to generate verifiable ML property cards using ML property attestation from TEEs.
TEEs already have the notion of hardware-assisted \emph{remote attestation} to prove local system or software configuration to a remote verifier. The trusted computing research community has extended this to the notion of \emph{property attestation}~\cite{propatt}, and recent developments by hardware vendors~\cite{amx,h100}, make it possible to train and run ML models efficiently inside TEEs. 
We are the first to propose a software framework for \emph{hardware-assisted ML property attestation}. 

\method establishes bindings between key components: 
\begin{enumerate*}[itemjoin={,\xspace},label=(\roman*)]
    \item the model and its inputs/outputs
    \item  the model and its accuracy with respect to a test dataset
    \item the model and its training dataset and configuration
    \item and distributional properties of a dataset.
\end{enumerate*} 
Bindings are signed with the TEE's secret attestation key, producing \emph{attestations}. Since these can be generated and validated by anyone with the requisite hardware, this approach is inherently scalable and allows verifiers to draw conclusions about the model and training dataset properties.
We implement the different attestations described in Section~\ref{sec:attestations}.
We revisit the completeness and how to extend \method for new attestations in Section~\ref{sec:discussions}.
We now discuss the design and implementation of \method for \prover and \verifier.

\input{figures/fig_implementation}

We present an overview of \method's implementation in Figure~\ref{fig:implementation}. A TEE runs a Python script (``measurer'') that measures properties of a model, dataset, or inference, and produces an attestation of a property card fragment (i.e.~a chunk of JSON with the same structure as model card metadata that will be combined with other such fragments to produce the full property card).
The combination of certificates and attestations is called the \emph{assertion bundle}. This approach can be made highly \emph{versatile}, as new enclaves can be endorsed without updating all verifiers' software.

We also consider assertions by trusted certifiers (e.g., CIFAR, Huggingface) who do not take advantage of remote attestation, but who can be trusted to provide external certificates asserting that some publicly available models or datasets are of good quality, or simply to name a dataset.
During verification, the assertion bundle can be passed to \verifier who verifies the attestations using certificates from the TEE manufacturer, and combines them to obtain a set of verified ML property cards.


We implement our proof of concept using Intel's SGX as a TEE platform, but the design is not dependent on platform-specific details. It can be equally implemented on any other TEE with remote attestation support. Different implementations can interoperate as long as the verifier can validate each TEE's remote attestations.

\subsection{The Prover}\label{sec:prover}

We implement \method's prover using Gramine~\cite{kuvaiskii2022computation}, which allows unmodified applications to run inside an SGX enclave. This allows \method to generate attestations for properties computed with off-the-shelf libraries like PyTorch~\cite{paszke2019pytorch}.

We implement the attestations from Section~\ref{sec:attestations}. The different inputs and outputs from different attestation enclaves are shown in Figure~\ref{fig:enclaves}:
\begin{enumerate}[leftmargin=*]
\item the \emph{dataset enclave} takes $\dtrain$ as an input and generates the distribution along with \Cref{att:propatt}.
\item the \emph{training enclave} takes $\dtrain$, $\modelarch$, and $\trainconfig$ as input, to train a model inside the enclave, and output the trained model $\model$ along with PoT~(\Cref{att:prooftr}).
\item the \emph{metric enclave} takes $\model$ and $\dtest$ as input, generating metrics for accuracy, fairness, and robustness, along with their attestations as an output (\Cref{att:accatt}, \Cref{att:fairatt}, and \Cref{att:robatt} respectively). 
\item the \emph{inference enclave} runs $\model$ after deployment, producing the model's output and corresponding IOAtt for each input (\Cref{att:inpmodout}).
\end{enumerate}

\input{figures/fig_enclaves}

The attestations are generated from a custom ML library built atop PyTorch. 
A hash of the Python code responsible for these attestations, along with PyTorch and all other files used by the enclave, is embedded into the enclave using Gramine, which provides a library OS that allows the integrity of the Python code to be verified when read from disk, incorporating it into the \texttt{MRENCLAVE} value included in SGX attestations.

\method includes a wrapper for Python's \texttt{open} library call, which is used to read input files into memory, hashing them at the same time, allowing PyTorch to randomly access measured files without risk of time-of-check-time-of-use (TOCTOU) vulnerabilities. We implement an additional mechanism to check the resulting hash against a manifest, allowing for a single short measurement to represent a large directory of input files without the need to keep them all in memory.

After the computation is complete, a hash of the output is incorporated into the user data field of an SGX report, which is then used to obtain a DCAP quote~\cite{dcap} that can be verified by remote parties.  This output takes the form of a JSON string representing a fragment of the model card metadata\footnote{\url{https://github.com/huggingface/hub-docs/blob/main/modelcard.md}}, where a model is named with the hash of its file.

\subsection{The Verifier}

\method's verifier takes two main inputs:
\begin{itemize}[leftmargin=*]
    \item A set of serialized attestations in JSON format, each containing an SGX quote---representing the attestation signature from \Cref{sec:attestations}---and base64-encoded output.
    \item A set of certifications relating SGX enclave identities to the property that they attest; in our case, a hard-coded mapping from SGX enclave measurements to JSON templates for the property card metadata that they are allowed to assert.
\end{itemize}
The attestations are designed in such a way that they directly map to YAML-formatted property card metadata, similar to that included in existing Model Cards. 
This greatly simplifies the verifier, which needs only to check that the attested JSON corresponds to the properties included in the corresponding certification.

\verifier checks each quote by validating it against the output of the corresponding attestation. \verifier then uses the enclave measurement from the quote to look up the associated certification, which contains a JSON structure that acts as a template for the payload. This template is then validated according to the following rules:
\begin{itemize}[leftmargin=*]
    \item \texttt{null} in the certification matches anything in the payload,
    \item The dictionaries in the certification correspond to dictionaries (or arrays of dictionaries) in the payload. The keys in the payload must match those in the certification, and the values in the payload must match the values in the certification,
    \item Strings, booleans, and integers in the certification match identical values (or arrays of identical values) in the payload,
    \item Other values are not allowed in the certification.
\end{itemize}
Specifically, we do not allow the use of floating-point values in order to avoid rounding issues due to verification; these must be represented as JSON strings.

For each model referenced by an attestation, \verifier produces a model card, formed by taking the union of the various model card metadata fragments to produce a single JSON dictionary that is reformatted into the YAML contents of the model card \texttt{model-index} field, which contains experimental data in a structured form.
Also, \method lets \verifier combine TEE attestations with other guarantees (e.g., external certificates in Section~\ref{sec:chain}). Assurances from ZKP-like mechanisms can be integrated into \method similarly.

\subsection{Authorization}

Enclaves should only generate attestations for certified properties, and the verifier should not trust attestations from unknown enclaves or those not certified to make a specific assertion.
To check for this, an endorser can provide an external certificate for an enclave, in addition to those for datasets or models, to indicate which properties each enclave is allowed to generate. 
These external certificates can be included in the assertion bundle and validated during the verification, though in our proof-of-concept they are hardcoded into the verifier.

%% file: figures/fig_implementation.tex
\begin{figure}[!htb]
\begin{center}
\resizebox{0.9\columnwidth}{!}{ 
\begin{tikzpicture}[line width=1pt]

\node (script) [draw, align=center, fill=green!8, minimum height=0.45cm, minimum width = 1.3cm] {\footnotesize Measurer};
\node (pytorch) [draw, right of = script, xshift=0.2cm, align=center] {\footnotesize Pytorch};
\node (python) [draw, below of=script, align=center, yshift=0.5cm, xshift=0.55cm, minimum width=2.4cm] {\footnotesize Python};
\node (gramine) [draw, below of=python, align=center, yshift=0.53cm, minimum width=2.4cm] {\footnotesize Gramine LibOS};

\begin{scope}[on background layer]
    \node (sgx1) [fit= (script) (pytorch) (python) (gramine), fill=black!8, rounded corners, draw=black, dashed, inner sep=.4cm, xshift=0.3cm, yshift=0.8cm,label={[shift={(0,-0.4)}]\footnotesize Trusted Certifier}] {};
\end{scope}

\begin{scope}[on background layer]
    \node (sgx2) [fit= (script) (pytorch) (python) (gramine), fill=orange!8, rounded corners, draw=black, dashed, inner sep=.4cm, xshift=0.15cm, yshift=0.45cm,label={[shift={(0,-0.4)}]\footnotesize SEV-SNP Virtual Machine}] {};
\end{scope}

\begin{scope}[on background layer]
    \node (sgx3) [fit=(script) (pytorch) (python) (gramine), fill=orange!8, rounded corners, draw=black, inner sep=.4cm, label={[shift={(0,-0.4)}]\footnotesize SGX Enclave}] {};
\end{scope}

\node (att) [draw, ellipse, below of =sgx3, yshift=-1cm, xshift=-1cm, fill= orange!8, align=center, minimum width=1.5cm] {\footnotesize Attestations};
\draw[->, ultra thick] ([xshift=-1cm]sgx3.south) -- (att.north);

\node (endorser) [draw, rounded corners, left of =sgx3, xshift=-3cm, yshift=0.75cm, fill=black!8, align=center, minimum width=3.2cm, minimum height=2.2cm] {\footnotesize Trusted Certifier};
\node (endorsements) [draw,ellipse, below of=endorser, yshift=-1.75cm, fill= black!8, align=center, minimum width=1.5cm] {\footnotesize Non-computational\\\footnotesize Property Certificates};
\draw[->, ultra thick, dashed] (endorser.south) -- (endorsements.north);

\node (bundle) [ellipse, draw, align=center, fill= orange!8, below of=endorser, xshift=1.5cm, yshift=-3cm] {\footnotesize Assertion\\\footnotesize Bundle};
\draw[->, ultra thick, dashed] (endorsements.south) -- ([yshift=0.2cm]bundle.west);
\draw[->, ultra thick] (att.south) -- ([yshift=0.2cm]bundle.east);


\node (verifier) [draw, rounded corners, below of =bundle, xshift=0.5cm, yshift=-0.75cm, fill=orange!8, align=center, minimum width=7cm, minimum height=1.25cm,label={[xshift=-3cm, yshift=-0.85cm]\footnotesize Verifier}] {};

\node (propcard) [draw, ellipse, fill=blue!8, below of =verifier, xshift=-0.5cm, yshift=-0.75cm, align=center, minimum width=4cm] {\footnotesize Property Cards};
\draw[->, ultra thick, dashed] ([xshift=-0.1cm]bundle.south) -- ([xshift=-0.1cm]propcard.north);
\draw[->, ultra thick] ([xshift=0.2cm]bundle.south) -- ([xshift=0.2cm]propcard.north);

\end{tikzpicture}
}
\end{center}
\caption{\textbf{Overview of \method's design:} Components already existing or part of the infrastructure are indicated in \colorbox{black!8}{gray} while components specific to \method are indicated in \colorbox{orange!8}{orange}. There can be different enclaves for different attestations which are generated by changing the ``measurer'' (in \colorbox{green!8}{green}). Dashed lines correspond to assertions by a trusted certifier, while solid lines correspond to assertions by a TEE.}
\label{fig:implementation}
\end{figure}

%% file: figures/fig_enclaves.tex
\begin{figure}[!htb]
\begin{center}
\begin{tikzpicture}[line width=1pt]

\node (trdata) [align=center, minimum width=1cm, draw, dashed] {\footnotesize Training Dataset\\\footnotesize ($\dtrainprov$)};
\node (config) [align=center, minimum width=1cm, draw, dashed, right of=trdata, xshift=2cm] {\footnotesize Training and Model\\\footnotesize Configuration ($\trainconfig$+$\modelarch$)};

\node (dataenclave) [draw, below of=trdata, align=center, yshift=-0.5cm, minimum width=2.5cm, minimum height=1cm, fill=orange!8, xshift=-2cm] {\footnotesize Dataset Enclave};
\node (script) [draw, align=center, below of=dataenclave, fill=green!8, minimum height=0.2cm, minimum width=1cm, xshift=0.75cm, yshift=0.65cm] {\footnotesize ....};
\node (trenclave) [draw, below of=trdata, align=center, yshift=-0.5cm, minimum width=2.5cm, minimum height=1cm, fill=orange!8, xshift=3cm] {\footnotesize Training Enclave};
\node (script) [draw, align=center, below of=trenclave, fill=green!8, minimum height=0.2cm, minimum width=1cm, xshift=0.75cm, yshift=0.65cm] {\footnotesize ....};
\draw[->, ultra thick] ([yshift=-0.5cm]trdata.west) -- (dataenclave.north);
\draw[->, ultra thick] ([yshift=-0.5cm]trdata.east) -- ([xshift=-1.2cm]trenclave.north);
\draw[->, ultra thick] (config.south) -- (trenclave.north);

\node (prop) [align=center, below of=dataenclave, yshift=-0.5cm, minimum width=1cm, draw, dashed, xshift=-0.9cm, fill=blue!8] {\footnotesize Distribution};
\node (distatt) [align=center, ellipse, below of=dataenclave, yshift=-0.5cm, minimum width=1cm, draw, dashed, xshift=0.9cm, fill=orange!8] {\footnotesize DistAtt};
\draw[->, ultra thick] (dataenclave.south) -- (prop.north);
\draw[->, ultra thick] (dataenclave.south) -- (distatt.north);

\node (model) [align=center, below of=trenclave, yshift=-0.5cm, minimum width=1cm, draw, dashed, xshift=-1.5cm, fill=blue!8] {\footnotesize Model};
\node (pot) [align=center, ellipse, below of=trenclave, yshift=-0.5cm, minimum width=1cm, draw, dashed, xshift=-0.2cm, fill=orange!8] {\footnotesize PoT};
\node (inp) [align=center, right of=pot, minimum width=1cm, draw, dashed, xshift=0.2cm] {\footnotesize Input};
\draw[->, ultra thick] (trenclave.south) -- (model.north);
\draw[->, ultra thick] (trenclave.south) -- (pot.north);

\node (infenclave) [draw, below of=trenclave, align=center, yshift=-2cm, minimum width=2.5cm, minimum height=1cm, fill=orange!8] {\footnotesize Inference Enclave};
\node (script) [draw, align=center, below of=infenclave, fill=green!8, minimum height=0.2cm, minimum width=1cm, xshift=0.75cm, yshift=0.65cm] {\footnotesize ....};
\node (metricenclave) [draw, below of=dataenclave, align=center, yshift=-2cm, minimum width=2.5cm, minimum height=1cm, fill=orange!8, xshift=1.75cm] {\footnotesize Metric Enclave};
\node (script) [draw, align=center, below of=metricenclave, fill=green!8, minimum height=0.2cm, minimum width=1cm, xshift=0.75cm, yshift=0.65cm] {\footnotesize ....};
\node (tedata) [align=center, left of=metricenclave, xshift=-1.6cm, minimum width=1cm, draw, dashed] {\footnotesize Test Dataset\\\footnotesize ($\dtest$)};

\draw[->, ultra thick] (inp.south) -- ([xshift=0.1cm]infenclave.north);
\draw[->, ultra thick] (model.south) -- (infenclave.north);
\draw[->, ultra thick] (model.south) -- (metricenclave.north);
\draw[->, ultra thick] (tedata.east) -- (metricenclave.west);

\node (metrics) [align=center, below of=metricenclave, yshift=-0.5cm, minimum width=1cm, draw, dashed, xshift=-2.75cm, fill=blue!8] {\footnotesize Metric};
\node (accatt) [align=center, ellipse, below of=metricenclave, yshift=-0.5cm, minimum width=1cm, draw, dashed, xshift=-1cm, fill=orange!8] {\footnotesize AccAtt};
\node (fairatt) [align=center, ellipse, right of=accatt, minimum width=1cm, draw, dashed, fill=orange!8,xshift=0.6cm] {\footnotesize FairAtt};
\node (robatt) [align=center, ellipse, below of=fairatt, minimum width=1cm, draw, dashed, fill=orange!8,xshift=-0.75cm, yshift=0.3cm] {\footnotesize RobAtt};
\begin{scope}[on background layer]
    \node (atts) [fit=(accatt) (fairatt) (robatt), fill=orange!8, rounded corners, draw=black, inner sep=.05cm, dashed] {};
\end{scope}
\draw[->, ultra thick] (metricenclave.south) -- (atts.north);
\draw[->, ultra thick] (metricenclave.south) -- (metrics.north);

\node (out) [align=center, below of=infenclave, yshift=-0.5cm, minimum width=1cm, draw, dashed, xshift=-0.75cm, fill=blue!8] {\footnotesize Output};
\node (ioatt) [align=center, ellipse, below of=infenclave, yshift=-0.5cm, minimum width=1cm, draw, dashed, xshift=0.6cm, fill=orange!8] {\footnotesize IOAtt};
\draw[->, ultra thick] (infenclave.south) -- (out.north);
\draw[->, ultra thick] (infenclave.south) -- (ioatt.north);

\end{tikzpicture}
\end{center}
\caption{\textbf{Inputs and outputs for different attestation enclaves.} \method includes four different enclaves which are indicated in \colorbox{orange!8}{orange}. Attestations are in ellipse, outputs are in \colorbox{blue!8}{blue}, and inputs to enclaves are \colorbox{white!8}{white}. We indicate measurer in \colorbox{green!8}{green}.}
\label{fig:enclaves}
\end{figure}

%% file: 6evaluation.tex
\section{Evaluation}\label{sec:evaluation}

We now evaluate \method and our proof-of-concept against the requirements identified in \Cref{sec:problem}.

\subsection{\ref{efficient}: Efficiency}

We assess efficiency by measuring the overhead from measurement and attestation in a confidential ML workload, exploring how it varies with model complexity and dataset size. We aim to validate the following conjectures: the overhead of \method is
\begin{enumerate}[leftmargin=*]
    \item determined by a roughly constant attestation overhead, and a measurement overhead dominated by the size of its inputs and outputs. We indicate the inputs and outputs for measurements across the four attestation types below:
        \begin{itemize}
            \item \textit{Distribution attestation:} Training dataset as input, and distribution value as output.
            \item \textit{Proof of training:} Training dataset as input, and trained model as output.
            \item \textit{Accuracy attestation:} Test dataset and trained model as input, and accuracy as output.
            \item \textit{Input-model-output attestation:} Trained model and data record for inference as input, and predictions as output.
        \end{itemize}
    \item small in comparison with the computational task being attested for all attestations except for IO attestation which we discuss separately in Section~\ref{sec:performance} (Input-Model-Output Attestation).
\end{enumerate}

\subsubsection{Experimental Setup}\label{sec:setup}

We now present hardware configuration, datasets, model architecture and training configurations, and metrics for evaluation. Reported values are averages of five runs.

\noindent\textbf{Hardware Configuration.}  We conduct an evaluation on a machine with the following components:
\begin{itemize}[leftmargin=*]
    \item \textbf{CPU:} Intel Xeon Silver 4514Y.
    \item \textbf{RAM:} 512GB, 8x 64GB (Micron Technology MTC40F2046S1RC48BA1 MHCC), 4800 MHz.
    \item \textbf{Storage:} Micron 7450 MTFDKBG1T9TFR, 1920GB
    \item \textbf{Mainboard:} Supermicro X13SEDW-F
\end{itemize}

\noindent\textbf{Datasets.} We consider three datasets of varying size drawn from several domains: a tabular dataset CENSUS~\cite{census}, an image dataset UTKFACE~\cite{utkface}, and a textual dataset IMDB~\cite{imdb}.
These datasets are standard benchmark used in ML literature. 

\noindent\textit{CENSUS} contains 48,842 records with 42 attributes from the 1994 US Census, including sensitive attributes like sex and race. The task is to predict whether an individual’s annual income exceeds 50,000 USD. A 50-50 train-test split yields 24,000 records each for training and testing. This translates to 2MB disk size for train and test dataset.

\noindent\textit{UTKFACE} is a large-scale face dataset with 48x48 RGB images for classifying individuals as young (<30 years). It includes 11,852 training and 11,853 testing samples, with sensitive attributes of race and sex.
This translates to 100MB disk size for train and test dataset.
\noindent\textit{IMDB} is a text dataset containing movie reviews from the IMDB website for binary classification of positive/negative sentiments. We use 37,500 training and 12,500 test data records. This translates to 49.5MB and 16.5MB disk size for train and test dataset respectively.

\noindent\textbf{Model Architecture.} We train two models on each dataset: a smaller model (tagged with the suffix ``-S''), and a larger model (tagged with the suffix ``-S'').
We summarize the models and the number of parameters in Table~\ref{tab:modelSummary}.

\begin{table}[!htb]
    \centering
    \scriptsize
    \caption{Summary of models across different datasets.}
    \begin{tabular}{l|c|c|c}
    \bottomrule

    \toprule
        \textbf{Model} & \textbf{Description} & \textbf{\#Parameters} & \textbf{Model Size (MB)} \\
        \midrule
        \textbf{CENSUS-S} & MLP: [128] & 12,290 & 0.05 \\
         \textbf{CENSUS-L} & MLP: [128, 256, 512, 256] & 308,482 & 1.2\\
        \midrule
        \textbf{UTKFACE-S} & VGG11 & 9,227,010 & 36.95\\
         \textbf{UTKFACE-L} & VGG16 & 14,724,162 & 58.96\\
        \midrule
        \textbf{IMDB-S} & LSTM: [64, 256, 256] & 920,385 & 3.69\\
        \textbf{IMDB-L} &LSTM: [64, 256, 256, 256, 256] & 1,973,057 & 7.60\\
    \bottomrule

    \toprule
    \end{tabular}
    \label{tab:modelSummary}
\end{table}

\noindent\textit{CENSUS-S} and \textit{CENSUS-L} are multi-layer perceptron (MLP) models with hidden layers of dimensions [128] and [128, 256, 512, 256] respectively with $\tanh$ activation functions. \textit{CENSUS-S} has $\sim$12K model parameters (0.05 MB) while \textit{CENSUS-L} has $\sim$308K parameters (1.2 MB). We train these models for 5 epochs which gives an accuracy of 78.33\% (\textit{CENSUS-S}) and 81.09\% (\textit{CENSUS-L}).

\noindent\textit{UTKFACE-S} and \textit{UTKFACE-L} are VGG11 and VGG13 models~\cite{vgg} respectively.
VGG11 with hidden layer dimensions: [64, `M', 128, `M', 256, 256, `M', 512, 512, `M', 512, 512, `M'] and VGG16 includes [64, 64, `M', 128, 128, `M', 256, 256, 256, `M', 512, 512, 512, `M', 512, 512, 512, `M'], where `M' indicates maxpool. \textit{UTKFACE-S} has $\sim$9.22 million parameters (36.95MB) and \textit{UTKFACE-L} with $\sim$14.72 million parameters (58.96MB). We train the models for 10 epochs to get 80.28\% (\textit{UTKFACE-S}) and 80.81\% (\textit{UTKFACE-L}) accuracy.

\noindent\textit{IMDB-S} and \textit{IMDB-L} are LSTM models with hidden layer dimensions of [64, 256, 256] and [64, 256, 256, 256, 256] respectively. \textit{IMDB-S} has $\sim$920K parameters (3.69MB) and \textit{IMDB-L} has 1.97 million parameters (7.60MB). We train these models for five epochs to get an accuracy of 85.60\% (\textit{IMDB-S}) and 86.18\% (\textit{IMDB-L}).






\noindent\textbf{Metrics.} We split the tasks into the following phases of computation, with \method's overhead marked in bold:
\begin{itemize}[leftmargin=*]
    \item \emph{Input loading} is the time spent reading the operation's input data.
    \item \emph{\textbf{Input measurement}} is the time spent hashing the input data to produce an input measurement.
    \item \emph{Data pre-processing} is the time spent on transforming the inputs (e.g., feature selection, data augmentation).
    \item \emph{Computation} is the time spent on the inference, training, or property measurement.
    \item \emph{\textbf{Output measurement}} is the time spent hashing the output data to produce an output measurement.
    \item \emph{Output storage} is the time spent formatting and writing the output data.
    \item \emph{\textbf{Attestation}} is the time spent producing the attestation of the measurements.
\end{itemize} 
\method's overhead is therefore the total of \emph{\textbf{input measurement}}, \emph{\textbf{output measurement}}, and \emph{\textbf{attestation}}. We use the total time spent on the remaining phases as the \emph{baseline}, which includes \emph{input loading}, \emph{data pre-processing}, \emph{computation}, and \emph{output storage}, against which we compare \method's overhead.

\subsubsection{Results}\label{sec:performance}

We present the overheads of attesting four types of ML properties for datasheets, model cards, and inference cards. 

\noindent\textbf{\underline{Attestations for Datasheets.}} We measure the execution time of distributional property attestation for verifiable datasheets.

\noindent\textbf{Distribution Attestation.} Table~\ref{tab:distAtt} summarizes the evaluation of distributional properties for CENSUS and UTKFACE, which include race and sex as sensitive attributes. We do not evaluate on the IMDB dataset since its text data does not contain any sensitive attributes.
We generate attestations for two types of distributional properties: 
\begin{enumerate*}[label=\roman*),itemjoin={,\xspace}]
\item the distribution of a sensitive attribute $z$ and
\item the distribution of $z$ given the classification label $y$ (denoted $z|y$ in Table~\ref{tab:distAtt}).
\end{enumerate*}

\input{tables/tab_distribution_new}


Attestation time remains nearly constant across datasets. As expected, the input measurement time is higher for the larger datasets (UTKFACE). The total overhead incurred by \method varies between 0.36\% and 2.05\%. The size of the output is negligible so there will be some minimal output measurement time which is incurred (0.72 ms for CENSUS and 0.87 ms for UTKFACE).

\noindent\textbf{\underline{Attestations for Model Cards.}} We report  the overhead for attesting two types of ML card properties: proof of training and accuracy.

\noindent\textbf{Proof of Training.} We train both small and large models on CENSUS, UTKFACE, and IMDB (Table~\ref{tab:training}). The highest recorded overhead was 0.32\% for the smallest model and dataset (CENSUS), while for larger datasets (UTKFACE and IMDB), \method’s overhead is negligible, at 0.01\% or lower. The attestation time does not increase, varying between 15ms and 31ms across each run, while the input measurement and output measurement times increase with the size of the input and output, as conjectured.

\input{tables/tab_training_new}

\noindent\textbf{Accuracy Attestation.} We measure the accuracy attestation performance for both small and large model variants across all datasets (Table~\ref{tab:accuracy}). The overhead of \method is less than 0.35\%. 
The input measurement overhead increases with input size, while the attestation time varies between approximately between 22ms and 27ms, consistent with the other attestations. 
The size of the output is negligible so there will be some minimal output measurement time which is incurred, where the overhead for -S and -L models are close to each other.
We note that there is a higher attestation time for -L models compared to -S models for CENSUS and UTKFACE. 
Since attestation is independent of the model size, we attribute this to some randomness during attestation.

\input{tables/tab_accuracy_new}

\noindent\textbf{\underline{Attestations for Inference Cards.}}
We measure the performance of input-model-output attestation for verifiable inference cards.

\noindent\textbf{Input-Model-Output Attestation.} We measure the execution time of input-model-output attestation for a single inference averaged over 100 runs (Table~\ref{tab:IO}). Since the size of the output is negligible the time for output measurement time is very small. Unlike the types of operations we have considered above, the baseline cost for a \emph{single} inference is also very small. Consequently, it is no surprise that although \method's computation time overhead for the attestation itself remains small (as in previous cases), the proportional overhead is relatively high, (between 39\% and 3955\%). The high overhead stems from the attestation signature being much more costly than inference computation, as shown in Table~\ref{tab:IO}.

\input{tables/tab_inout_new}

We describe how to reduce the per-inference overhead.
One approach is to batch multiple inputs from a single client and attest them together, spreading the cost of a single attestation or signature across multiple tasks, resulting in overhead similar to accuracy attestation in Table~\ref{tab:accuracy}.
Another approach to reduce cost is to generate a signing keypair during TA initialization, perform a single attestation of the public key, and then sign each inference result for indirect, low-cost attestation.
We compare the time taken for signing with a keypair versus invoking hardware attestation.
We find that the resulting overhead computed using signing (``Overhead w/ sgn'' in Table~\ref{tab:accuracy}) is drastically less than using attestation (``Overhead w/ att'').


\noindent\textbf{\underline{Summary.}} These findings highlight \method's efficiency, with minimal overhead in comparison with the baseline cost of the computation being attested. The only exception to this is for inference attestation, where the time spent on inference is more comparable to the time spent on attestation; we suggest a mitigation for this by using signing key-pair in software instead of hardware-based attestation.
Verification costs are low, involving only certificate checking, and are independent of model size.
Hence, \method satisfies \ref{efficient}.

\subsection{\ref{scalable}: Scalability}
Mainstream TEEs provide attestations based on digital signatures that can be repeatedly verified by multiple verifiers. Additionally, an assertion bundle can include attestations from various provers, which are standardized into a common format. This allows \method to scale to large numbers of independent provers, without relying on a single entity or TEE type to generate all proofs needed for a verifiable property card.
Hence, \method satisfies \ref{scalable}.

\subsection{\ref{versatile}: Versatility}
\method's property definitions are software-based, with hardware only used for isolation and binary attestation, making it easy to extend to new properties. \prover{s} can write Python scripts to measure new properties, get them certified, and \verifier{s} can validate the claims. The attestation and certification formats map directly to YAML property card metadata, so \verifier requires no modification to support new metrics or ML property cards. Thus, \method satisfies \ref{versatile}.

%% file: tables/tab_distribution_new.tex

\setlength\tabcolsep{2pt}
\begin{table}[!htb]
    \caption{Performance of distributional property attestation. Operations in \colorbox{gray!15}{gray} form the baseline (\colorbox{orange!15}{orange}), while those in \colorbox{blue!15}{blue} form the \method's overhead (\colorbox{orange!15}{orange}). We omit "Output Size" as the output is the distribution value, which is a single value of negligible size.}
    \centering
    \scriptsize
    \begin{tabular}{l|c|c}
    \bottomrule

    \toprule
   \textbf{Dataset} & \textbf{CENSUS} & \textbf{UTKFACE} \\
    \midrule
\textbf{Preprocessing} & \cellcolor{gray!15}9.90 s & \cellcolor{gray!15}54.22 s\\
\textbf{Input Load} & \cellcolor{gray!15}28.89 ms & \cellcolor{gray!15}1.78 s\\
\textbf{Computation (Race: z)} & \cellcolor{gray!15}73.29 ms & \cellcolor{gray!15}66.28 ms\\
\textbf{Computation (Sex: z)} & \cellcolor{gray!15}4.38 ms & \cellcolor{gray!15}3.78 ms\\
\textbf{Computation (Race: z|y)} & \cellcolor{gray!15}26.35 ms & \cellcolor{gray!15}23.49 ms\\
\textbf{Computation (Sex: z|y)} & \cellcolor{gray!15}10.43 ms & \cellcolor{gray!15}8.74 ms\\
\textbf{Output Storage} & \cellcolor{gray!15}1.05 ms & \cellcolor{gray!15}1.11 ms\\
\midrule
\textbf{Baseline} & \cellcolor{orange!15}10.01 s & \cellcolor{orange!15}54.33 s\\
\midrule
\textbf{Input Meas.} & \cellcolor{blue!15}5.90 ms & \cellcolor{blue!15}1.10 s\\
Input Size & \cellcolor{blue!15}2 MB & \cellcolor{blue!15}100 MB\\
\textbf{Output Meas.} & \cellcolor{blue!15}0.72 ms & \cellcolor{blue!15}0.87 ms\\
\textbf{Attestation} & \cellcolor{blue!15}29.85 ms & \cellcolor{blue!15}31.23 ms\\
\midrule
\textbf{Overhead} & \cellcolor{orange!15}36.46 ms & \cellcolor{orange!15}1.14 s\\
\textbf{Overhead (\%)} & \cellcolor{orange!15}0.36 & \cellcolor{orange!15}2.05\\
    \bottomrule

    \toprule
    \end{tabular}
    \label{tab:distAtt}
\end{table}

%% file: tables/tab_training_new.tex

\begin{table}[!htb]
    \caption{Performance of proof of training. Operations in \colorbox{gray!15}{gray} form the baseline (\colorbox{orange!15}{orange}), while those in \colorbox{blue!15}{blue} form the \method's overhead (\colorbox{orange!15}{orange}).}
    \centering
    \footnotesize
    \resizebox{\columnwidth}{!}{%
    \begin{tabular}{l|c|c||c|c||c|c}
    \bottomrule

    \toprule
    \textbf{Dataset} & \multicolumn{2}{c||}{\textbf{CENSUS}}  & \multicolumn{2}{c||}{\textbf{UTKFACE}} & \multicolumn{2}{c}{\textbf{IMDB}} \\
    \textbf{Model variant} & \textbf{-S} & \textbf{-L}  & \textbf{-S} & \textbf{-L} & \textbf{-S} & \textbf{-L} \\
    \midrule
\textbf{Preprocessing} & \cellcolor{gray!15}9.95 s & \cellcolor{gray!15}9.71 s & \cellcolor{gray!15}54.62 s & \cellcolor{gray!15}54.42 s & \cellcolor{gray!15}70.31 s & \cellcolor{gray!15}70.35 s\\
\textbf{Input Load} & \cellcolor{gray!15}26.49 ms & \cellcolor{gray!15}28.05 ms & \cellcolor{gray!15}1.80 s & \cellcolor{gray!15}1.83 s & \cellcolor{gray!15}0.33 s & \cellcolor{gray!15}0.35 s\\
\textbf{Computation} & \cellcolor{gray!15}973.24 ms & \cellcolor{gray!15}0.03 min & \cellcolor{gray!15}310.11 min & \cellcolor{gray!15}643.13 min & \cellcolor{gray!15}1803.16 min & \cellcolor{gray!15}3563.11 min\\
\textbf{Output Storage} & \cellcolor{gray!15}0.74 ms & \cellcolor{gray!15}0.74 ms & \cellcolor{gray!15}1.81 ms & \cellcolor{gray!15}1.84 ms & \cellcolor{gray!15}0.97 ms & \cellcolor{gray!15}1.09 ms\\
\midrule
\textbf{Baseline} & \cellcolor{orange!15}10.95 s & \cellcolor{orange!15}11.53 s & \cellcolor{orange!15}311.05 min & \cellcolor{orange!15}644.06 min & \cellcolor{orange!15}1804.34 min & \cellcolor{orange!15}3564.29 min\\
\midrule
\textbf{Input Meas.} & \cellcolor{blue!15}5.86 ms & \cellcolor{blue!15}5.88 ms & \cellcolor{blue!15}1.13 s  & \cellcolor{blue!15}1.14 s & \cellcolor{blue!15}96.40 ms & \cellcolor{blue!15}97.35 ms\\
Input Size (MB) & \cellcolor{blue!15}2.00  & \cellcolor{blue!15}2.00  & \cellcolor{blue!15}100.00  & \cellcolor{blue!15}100.00  & \cellcolor{blue!15}49.50  & \cellcolor{blue!15}49.50 \\
\textbf{Output Meas.} (ms) & \cellcolor{blue!15}0.23 & \cellcolor{blue!15}1.12 & \cellcolor{blue!15}24.77 & \cellcolor{blue!15}226.79 & \cellcolor{blue!15}2.76 & \cellcolor{blue!15}5.47\\
Output Size (MB) & \cellcolor{blue!15}0.05  & \cellcolor{blue!15}1.20  & \cellcolor{blue!15}36.95  & \cellcolor{blue!15}58.96  & \cellcolor{blue!15}3.69  & \cellcolor{blue!15}7.60 \\
\textbf{Attestation} (ms) & \cellcolor{blue!15}29.16 & \cellcolor{blue!15}16.91 & \cellcolor{blue!15}25.82 & \cellcolor{blue!15}29.78 & \cellcolor{blue!15}26.66 & \cellcolor{blue!15}19.47\\
\midrule
\textbf{Overhead} & \cellcolor{orange!15}35.25 ms & \cellcolor{orange!15}23.91 ms & \cellcolor{orange!15}1.18 s & \cellcolor{orange!15}1.39 s & \cellcolor{orange!15}125.82 ms & \cellcolor{orange!15}122.30 ms\\
\textbf{Overhead (\%)} & \cellcolor{orange!15}0.32 & \cellcolor{orange!15}0.21 & \cellcolor{orange!15}0.01 & \cellcolor{orange!15}0.01 & \cellcolor{orange!15}0.00 & \cellcolor{orange!15}0.00\\
    \bottomrule

    \toprule
    \end{tabular}
    }
    \label{tab:training}
\end{table}

%% file: tables/tab_accuracy_new.tex

\begin{table}[!htb]
    \caption{Performance of accuracy attestation. Operations in \colorbox{gray!15}{gray} form the baseline (\colorbox{orange!15}{orange}), while those in \colorbox{blue!15}{blue} form the \method's overhead (\colorbox{orange!15}{orange}). We omit "Output Size" as the output is the accuracy, which is a single value of negligible size.}
    \centering
    \footnotesize
    \resizebox{\columnwidth}{!}{%
    \begin{tabular}{l|c|c||c|c||c|c}
    \bottomrule

    \toprule
    \textbf{Dataset} & \multicolumn{2}{c||}{\textbf{CENSUS}}  & \multicolumn{2}{c||}{\textbf{UTKFACE}} & \multicolumn{2}{c}{\textbf{IMDB}} \\
    \textbf{Model variant} & \textbf{-S} & \textbf{-L}  & \textbf{-S} & \textbf{-L} & \textbf{-S} & \textbf{-L} \\
    \midrule
\textbf{Preprocessing} & \cellcolor{gray!15}9.46 s & \cellcolor{gray!15}10.14 s & \cellcolor{gray!15}54.74 s & \cellcolor{gray!15}55.15 s & \cellcolor{gray!15}70.17 s & \cellcolor{gray!15}70.37 s\\
\textbf{Input Load} & \cellcolor{gray!15}28.15 ms & \cellcolor{gray!15}34.34 ms & \cellcolor{gray!15}1.96 s & \cellcolor{gray!15}2.08 s & \cellcolor{gray!15}0.36 s & \cellcolor{gray!15}0.37 s\\
\textbf{Computation} & \cellcolor{gray!15}228.62 ms & \cellcolor{gray!15}306.03 ms & \cellcolor{gray!15}10.45 min & \cellcolor{gray!15}21.38 min & \cellcolor{gray!15}53.78 min & \cellcolor{gray!15}107.46 min\\
\textbf{Output Storage} & \cellcolor{gray!15}0.95 ms & \cellcolor{gray!15}0.90 ms & \cellcolor{gray!15}1.60 ms & \cellcolor{gray!15}1.51 ms & \cellcolor{gray!15}1.70 ms & \cellcolor{gray!15}1.74 ms\\
\midrule
\textbf{Baseline} & \cellcolor{orange!15}9.72 s & \cellcolor{orange!15}10.48 s & \cellcolor{orange!15}11.40 min & \cellcolor{orange!15}22.33 min & \cellcolor{orange!15}54.95 min & \cellcolor{orange!15}108.64 min\\
\midrule
\textbf{Input Meas.} & \cellcolor{blue!15}7.03 ms & \cellcolor{blue!15}8.68 ms & \cellcolor{blue!15}1.15 s & \cellcolor{blue!15}1.24 s & \cellcolor{blue!15}0.10 s & \cellcolor{blue!15}0.11 s\\
Input Size (MB) & \cellcolor{blue!15}2.05 & \cellcolor{blue!15}3.20  & \cellcolor{blue!15}136.95  & \cellcolor{blue!15}158.96  & \cellcolor{blue!15}20.19  & \cellcolor{blue!15}24.10 \\
\textbf{Output Meas.} & \cellcolor{blue!15}0.01 ms & \cellcolor{blue!15}0.02 ms & \cellcolor{blue!15}0.40 ms & \cellcolor{blue!15}0.32 ms & \cellcolor{blue!15}0.44 ms & \cellcolor{blue!15}0.33 ms\\
\textbf{Attestation} & \cellcolor{blue!15}26.92 ms & \cellcolor{blue!15}24.15 ms & \cellcolor{blue!15}25.93 ms & \cellcolor{blue!15}22.19 ms & \cellcolor{blue!15}22.61 ms & \cellcolor{blue!15}27.20 ms\\
\midrule
\textbf{Overhead} & \cellcolor{orange!15}33.97 ms & \cellcolor{orange!15}32.85 ms & \cellcolor{orange!15}1.18 s & \cellcolor{orange!15}1.27 s & \cellcolor{orange!15}0.13 s & \cellcolor{orange!15}0.14 s\\
\textbf{Overhead (\%)} & \cellcolor{orange!15}0.35 & \cellcolor{orange!15}0.31 & \cellcolor{orange!15}0.17 & \cellcolor{orange!15}0.19 & \cellcolor{orange!15}0.00 & \cellcolor{orange!15}0.00\\
    \bottomrule

    \toprule
    \end{tabular}
    }
    \label{tab:accuracy}
\end{table}

%% file: tables/tab_inout_new.tex
\begin{table}[!htb]
    \caption{Performance of input-model-output attestation. Operations in \colorbox{gray!15}{gray} form the baseline (\colorbox{orange!15}{orange}), while those in \colorbox{blue!15}{blue} form the \method's overhead (\colorbox{orange!15}{orange}). We omit "Output Size" as the output is the prediction, which is a single value of negligible size. We report the overhead computed using attestation as ``w/ att'' and that with signing using keypair as ``w/ sgn''.}
    \centering
    \footnotesize
    \resizebox{\columnwidth}{!}{%
    \begin{tabular}{l|c|c||c|c||c|c}
    \bottomrule

    \toprule
    \textbf{Dataset} & \multicolumn{2}{c||}{\textbf{CENSUS}}  & \multicolumn{2}{c||}{\textbf{UTKFACE}} & \multicolumn{2}{c}{\textbf{IMDB}} \\
    \textbf{Model variant} & \textbf{-S} & \textbf{-L}  & \textbf{-S} & \textbf{-L} & \textbf{-S} & \textbf{-L} \\
    \midrule
\textbf{Input Load} & \cellcolor{gray!15}0.25 ms & \cellcolor{gray!15}6.13 ms & \cellcolor{gray!15}221.98 ms & \cellcolor{gray!15}336.42 ms & \cellcolor{gray!15}22.62 ms & \cellcolor{gray!15}41.47 ms\\
\textbf{Computation} & \cellcolor{gray!15}0.15 ms & \cellcolor{gray!15}0.40 ms & \cellcolor{gray!15}5.79 ms & \cellcolor{gray!15}14.05 ms & \cellcolor{gray!15}29.96 ms & \cellcolor{gray!15}57.97 ms\\
\textbf{Output Storage} & \cellcolor{gray!15}0.35 ms & \cellcolor{gray!15}1.33 ms & \cellcolor{gray!15}1.62 ms & \cellcolor{gray!15}1.83 ms & \cellcolor{gray!15}1.59 ms & \cellcolor{gray!15}1.61 ms\\
\midrule
\textbf{Baseline} & \cellcolor{orange!15}0.75 ms & \cellcolor{orange!15}7.87 ms & \cellcolor{orange!15}229.39 ms & \cellcolor{orange!15}352.29 ms & \cellcolor{orange!15}54.17 ms & \cellcolor{orange!15}101.05 ms\\
\midrule
\textbf{Input Meas.} & \cellcolor{blue!15}0.68 ms & \cellcolor{blue!15}2.38 ms & \cellcolor{blue!15}62.98 ms & \cellcolor{blue!15}96.70 ms & \cellcolor{blue!15}5.95 ms & \cellcolor{blue!15}11.97 ms\\
Input Size (MB) & \cellcolor{blue!15}0.05  & \cellcolor{blue!15}1.20  & \cellcolor{blue!15}36.95  & \cellcolor{blue!15}58.96  & \cellcolor{blue!15}3.69  & \cellcolor{blue!15}7.60 \\
\textbf{Output Meas.} & \cellcolor{blue!15}0.02 ms & \cellcolor{blue!15}0.40 ms & \cellcolor{blue!15}0.50 ms & \cellcolor{blue!15}0.56 ms & \cellcolor{blue!15}0.46 ms & \cellcolor{blue!15}0.60 ms\\
\textbf{Attestation} & \cellcolor{blue!15}28.97 ms & \cellcolor{blue!15}26.26 ms & \cellcolor{blue!15}27.38 ms & \cellcolor{blue!15}29.26 ms & \cellcolor{blue!15}23.13 ms & \cellcolor{blue!15}31.78 ms\\
\midrule
\textbf{Overhead w/ att} & \cellcolor{orange!15}29.67 ms & \cellcolor{orange!15}29.04 ms & \cellcolor{orange!15}90.86 ms & \cellcolor{orange!15}126.52 ms & \cellcolor{orange!15}29.54 ms & \cellcolor{orange!15}44.35 ms\\
\textbf{Overhead w/ att (\%)} & \cellcolor{orange!15}3955.35 & \cellcolor{orange!15}369.18 & \cellcolor{orange!15}39.61 & \cellcolor{orange!15}55.15 & \cellcolor{orange!15}54.52 & \cellcolor{orange!15}81.86\\
\midrule
\textbf{Signing in s/w} & \cellcolor{blue!15}0.18 ms & \cellcolor{blue!15}3.79 ms & \cellcolor{blue!15}4.40 ms & \cellcolor{blue!15}4.84 ms & \cellcolor{blue!15}4.64 ms & \cellcolor{blue!15}4.56 ms\\
\textbf{Overhead w/ sgn} & \cellcolor{orange!15}0.88 ms & \cellcolor{orange!15}6.56 ms & \cellcolor{orange!15}67.87 ms & \cellcolor{orange!15}102.10 ms & \cellcolor{orange!15}11.05 ms & \cellcolor{orange!15}17.13 ms\\
\textbf{Overhead w/ sgn (\%)} & \cellcolor{orange!15}1.17 & \cellcolor{orange!15}0.83 & \cellcolor{orange!15}0.30 & \cellcolor{orange!15}0.29 & \cellcolor{orange!15}0.20 & \cellcolor{orange!15}0.17 \\
    \bottomrule

    \toprule
    \end{tabular}
    }
    \label{tab:IO}
\end{table}

%% file: 7related.tex
\section{Related Work}\label{sec:related}

\noindent\textbf{ML-based Auditing.}
Several ML-based techniques, such as membership inference attacks, are used for auditing models by checking if data requested for removal under the right to be forgotten belongs to the training set~\cite{auditingMIA,miao2021audio}. 
Distribution inference attacks can also be used for inference-based property attestation~\cite{duddu2023attesting,juarez2022black}.
Both membership inference and distribution inference are unreliable and ineffective for auditing~\cite{rezaei2021difficulty,miaRepudiation,duddu2023attesting}.
Jia et al.~\cite{jia2021proof} propose proof of training to verify if a model was trained on a specific dataset. However, it can be easily evaded~\cite{PoLbroken,zhang2022adversarial}. Thus, ML-based techniques are unreliable for auditing.

\noindent\textbf{Certified ML.} Several works provide ``certificates'' as accuracy bounds in the presence of adversarial examples (e.g., Li et al.~\cite{li2023sok}). Differential privacy implementations may have bugs, leading to mismatched privacy bounds, prompting audits of such models~\cite{steinke2024privacy,jagielski2020auditing,nasr2023tight}. Additionally, bounds for group and individual fairness can be generated~\cite{jin2022input,khedr2023certifair,urban20}.
These bounds are generated using randomized smoothing or formal theory. This is an orthogonal problem as attestation attestation guarantees the computations done, not their correctness.
For instance, attestation of differential privacy training only confirms that differential privacy was used, not whether its implementation or guarantee is correct.

\noindent\textbf{Cryptographic Primitives for Attestation.} 
ZKPs have been widely used for verifiable inference to ensure model outputs are untampered~\cite{zkCNN,zenCompiler,Weng2021MystiqueEC,zkMLaaS,veriTrain,safetyNets}. ZKPs for training, though in early stages, have been used for proof of training~\cite{zkDL,shamsabadi2023confidentialprofitt,zkpTraining,zkpTraining2,eisenhofer2022verifiable}. However, generating proofs for each training epoch is inefficient, limiting this approach to simple ML models~\cite{shamsabadi2023confidentialprofitt,zkpTraining}. Extending these approaches to neural networks is an open problem due to significant overhead~\cite{zkDL,zkpTraining2}. Similarly, MPC for proof of training requires training the model for each attestation, incurring high overhead~\cite{duddu2023attesting}. Cryptographic primitives can also verify properties like fairness~\cite{shamsabadi2023confidentialprofitt,fairaudit1,fairaudit2,fairaudit3,yadav2024fairproof} and privacy~\cite{confidentialdpproof}.
Srivastava et al.~\cite{srivastava2024optimistic} propose a verifiable training scheme to ensure determinism, reduce hardware discrepancies, and improve auditing performance~\cite{zkpTraining}. MPC has also been used to verify the entire training and inference pipeline, but requires multiple interactions~\cite{lycklama2024holding}.

\noindent\textbf{Confidential Computing for ML.} TEEs have been widely explored for both training~\cite{mo2021ppfl} and inference of ML models~\cite{mo2020darknetz} (see survey by Mo et al.~\cite{mo2022sok}), often using optimizations like model partitioning. TEEs can also be combined with GPUs for efficient execution, adding cryptographic checks for data integrity between TEE and GPU~\cite{tramer2018slalom}. 
Prior TEE-based approaches focus on data confidentiality (e.g., private inference)~\cite{mo2022sok}, but ML verifiability requires integrity guarantees. We are not aware of any prior ML verifiability approaches using TEEs.

%% file: 8discussions.tex
\section{Discussion and Summary}\label{sec:discussions}

\noindent\textbf{Completeness.} We illustrate three types of ML property cards and four type of model attestations in Section~\ref{sec:attestations}, but do not claim completeness.
There could be other attestations types and ML property cards but \method can be extended to account for them by modifying the ``measurer'' script in the prover (Section~\ref{sec:prover}).
Any property that can be expressed in Python code can be attested using Laminator. 
For instance, \method can be extended to include other attestation types, such as ecosystem cards, a combination of multiple models trained on different datasets, using verifiable model cards and datasheets~\cite{bommasani2023ecosystem}.
Also, we can attest various steps in DPSGD training within our framework to demonstrate to a verifier that the prover has used the algorithm. 
Model explanations (during inference) will be handled like Input-Model-Output attestation.
While we focus on static properties for various ML operations, there are other runtime properties such as environmental impact of running ML models inside GPUs (e.g., carbon emission). We leave the design of attestations for such runtime properties for future.

\noindent\textbf{Security Analysis and Attacks on TEEs.} \method does not introduce any cryptographic protocols. Its security guarantees are inherited directly from the underlying attestation mechanism. 
Several prior works have demonstrated attacks on TEEs, particularly side-channel attacks, that can compromise confidentiality~\cite{sgxfail}. Attestation does not introduce new privacy risks, as no additional sensitive data is added to the TEE. \method relies on the integrity of the underlying TEE attestation mechanism to provide integrity guarantees for property attestation, complementing the confidentiality properties already offered by the TEE. Hence, leakage of model and training data via side channels is not a concern in our setting.
Leakage of attestation keys can compromise the integrity of the attestation mechanism~\cite{217543,sgxfail}, but this is orthogonal to our approach and not affected by \method's design. Mitigating such risks through side-channel-resistant code~\cite{constantine}, cache partitioning~\cite{sanctum}, or other techniques~\cite{stt} is an active area of research by TEE manufacturers.
Hence, \method relies on the platform developer to ensure attestation integrity, using hardware and software mechanisms to prevent subversion, such as side-channel or other attacks to extract attestation keys.

\noindent\textbf{Improving Efficiency with GPUs.} 
Our idea is independent of the specific TEE in which the computation is attested.
Most current TEEs are CPU-based, while most ML tasks use GPUs. New TEE-enabled GPUs like Nvidia's H100~\cite{h100} enable CPU-based TEEs to accelerate computations, allowing evaluation of larger models, including language models, with comparable performance~\cite{H100-performance}.
We plan to extend support to H100 GPUs.
CPU extensions like Intel AMX~\cite{amx} can accelerate ML operations within existing TEEs, enabling faster inference and training than traditional CPU implementations.
Where the performance of CPUs are insufficient, hybrid approaches like Slalom~\cite{tramer2018slalom} and TEESlice~\cite{noprivacy} enable a CPU-based TEE to offload model layers to a GPU, using cryptography to ensure data confidentiality and output integrity. This speeds up computation compared to CPU-only execution, but the added cryptographic overhead and data transfer between CPU and GPU reduce performance compared to TEE-enabled GPUs.


\noindent\textbf{Summary.} ML property cards, widely used by companies like Huggingface, require verification to ensure their correctness. We propose \method, the first framework for generating verifiable ML property cards using hardware-assisted attestations. We demonstrate that \method is efficient, scalable, and versatile, enabling higher-performance, verifiably trustworthy ML-based services.








%% file: 00ack.tex
\section*{Acknowledgments}
This work is supported in part by the Research Council of Finland (Decision 339514), Intel Labs, the Government of Ontario, and the Natural Sciences and Engineering Research Council of Canada (grant number RGPIN-2020-04744).
Vasisht is supported by IBM PhD Fellowship, David R. Cheriton Scholarship, and Cybersecurity and Privacy Excellence Graduate Scholarship.
The views expressed in the paper are those of the authors and do not reflect the position of the funding agencies.
%
%
The authors thank Prach Chantasantitam and Asim Waheed for their help running the code.

%% file: 0arxiv.bbl

\begin{thebibliography}{83}


\ifx \showCODEN    \undefined \def \showCODEN     #1{\unskip}     \fi
\ifx \showDOI      \undefined \def \showDOI       #1{#1}\fi
\ifx \showISBNx    \undefined \def \showISBNx     #1{\unskip}     \fi
\ifx \showISBNxiii \undefined \def \showISBNxiii  #1{\unskip}     \fi
\ifx \showISSN     \undefined \def \showISSN      #1{\unskip}     \fi
\ifx \showLCCN     \undefined \def \showLCCN      #1{\unskip}     \fi
\ifx \shownote     \undefined \def \shownote      #1{#1}          \fi
\ifx \showarticletitle \undefined \def \showarticletitle #1{#1}   \fi
\ifx \showURL      \undefined \def \showURL       {\relax}        \fi
\providecommand\bibfield[2]{#2}
\providecommand\bibinfo[2]{#2}
\providecommand\natexlab[1]{#1}
\providecommand\showeprint[2][]{arXiv:#2}

\bibitem[Abadi et~al\mbox{.}(2016)]%
        {abadi2016deep}
\bibfield{author}{\bibinfo{person}{Martin Abadi} {et~al\mbox{.}}}
  \bibinfo{year}{2016}\natexlab{}.
\newblock \showarticletitle{Deep Learning with Differential Privacy}. In
  \bibinfo{booktitle}{\emph{CCS}} (Vienna, Austria).
  \bibinfo{publisher}{Association for Computing Machinery},
  \bibinfo{address}{New York, NY, USA}, \bibinfo{pages}{308–318}.
\newblock
\showISBNx{9781450341394}
\urldef\tempurl%
\url{https://doi.org/10.1145/2976749.2978318}
\showDOI{\tempurl}


\bibitem[Abbaszadeh et~al\mbox{.}(2024)]%
        {zkpTraining2}
\bibfield{author}{\bibinfo{person}{Kasra Abbaszadeh} {et~al\mbox{.}}}
  \bibinfo{year}{2024}\natexlab{}.
\newblock \bibinfo{title}{Zero-Knowledge Proofs of Training for Deep Neural
  Networks}.
\newblock \bibinfo{howpublished}{Cryptology ePrint Archive, Paper 2024/162}.
\newblock
\urldef\tempurl%
\url{https://eprint.iacr.org/2024/162}
\showURL{%
\tempurl}
\newblock
\shownote{\url{https://eprint.iacr.org/2024/162}}.


\bibitem[Bommasani et~al\mbox{.}(2023)]%
        {bommasani2023ecosystem}
\bibfield{author}{\bibinfo{person}{Rishi Bommasani} {et~al\mbox{.}}}
  \bibinfo{year}{2023}\natexlab{}.
\newblock \showarticletitle{Ecosystem graphs: The social footprint of
  foundation models}.
\newblock \bibinfo{journal}{\emph{arXiv preprint arXiv:2303.15772}}
  (\bibinfo{year}{2023}).
\newblock


\bibitem[Borrello et~al\mbox{.}(2021)]%
        {constantine}
\bibfield{author}{\bibinfo{person}{Pietro Borrello} {et~al\mbox{.}}}
  \bibinfo{year}{2021}\natexlab{}.
\newblock \showarticletitle{Constantine: Automatic Side-Channel Resistance
  Using Efficient Control and Data Flow Linearization}. In
  \bibinfo{booktitle}{\emph{CCS}}. \bibinfo{publisher}{{ACM}},
  \bibinfo{pages}{715--733}.
\newblock
\urldef\tempurl%
\url{https://doi.org/10.1145/3460120.3484583}
\showDOI{\tempurl}


\bibitem[Bulck et~al\mbox{.}(2018)]%
        {217543}
\bibfield{author}{\bibinfo{person}{Jo~Van Bulck} {et~al\mbox{.}}}
  \bibinfo{year}{2018}\natexlab{}.
\newblock \showarticletitle{Foreshadow: Extracting the Keys to the {Intel SGX}
  Kingdom with Transient Out-of-Order Execution}. In
  \bibinfo{booktitle}{\emph{USENIX Sec}}. \bibinfo{publisher}{USENIX
  Association}, \bibinfo{address}{Baltimore, MD},
  \bibinfo{pages}{991{\textendash}1008}.
\newblock
\showISBNx{978-1-939133-04-5}
\urldef\tempurl%
\url{https://www.usenix.org/conference/usenixsecurity18/presentation/bulck}
\showURL{%
\tempurl}


\bibitem[Chang et~al\mbox{.}(2023)]%
        {holmes}
\bibfield{author}{\bibinfo{person}{Ian Chang} {et~al\mbox{.}}}
  \bibinfo{year}{2023}\natexlab{}.
\newblock \showarticletitle{{HOLMES}: Efficient Distribution Testing for Secure
  Collaborative Learning}. In \bibinfo{booktitle}{\emph{USENIX Sec}}.
  \bibinfo{publisher}{USENIX Association}, \bibinfo{address}{Anaheim, CA},
  \bibinfo{pages}{4823--4840}.
\newblock
\showISBNx{978-1-939133-37-3}
\urldef\tempurl%
\url{https://www.usenix.org/conference/usenixsecurity23/presentation/chang}
\showURL{%
\tempurl}


\bibitem[Christian(2021)]%
        {christian2021alignment}
\bibfield{author}{\bibinfo{person}{Brian Christian}.}
  \bibinfo{year}{2021}\natexlab{}.
\newblock \bibinfo{booktitle}{\emph{The Alignment Problem: How Can Machines
  Learn Human Values?}}
\newblock \bibinfo{publisher}{Atlantic Books}.
\newblock


\bibitem[Commission(2024)]%
        {eur}
\bibfield{author}{\bibinfo{person}{European Commission}.}
  \bibinfo{year}{2024}\natexlab{}.
\newblock \showarticletitle{{Regulation (EU) 2024/1689 (Artificial Intelligence
  Act)}}.
\newblock \bibinfo{journal}{\emph{https://artificialintelligenceact.eu/}}
  (\bibinfo{year}{2024}).
\newblock


\bibitem[Congress(2022)]%
        {congress}
\bibfield{author}{\bibinfo{person}{U.S. Congress}.}
  \bibinfo{year}{2022}\natexlab{}.
\newblock \showarticletitle{{H.R.~6580 -- Algorithmic Accountability Act}}.
\newblock
  \bibinfo{journal}{\emph{https://www.congress.gov/bill/117th-congress/house-bill/6580/text}}
  (\bibinfo{year}{2022}).
\newblock


\bibitem[Costan et~al\mbox{.}(2016)]%
        {sanctum}
\bibfield{author}{\bibinfo{person}{Victor Costan} {et~al\mbox{.}}}
  \bibinfo{year}{2016}\natexlab{}.
\newblock \showarticletitle{Sanctum: Minimal Hardware Extensions for Strong
  Software Isolation}. In \bibinfo{booktitle}{\emph{Usenix Sec}},
  \bibfield{editor}{\bibinfo{person}{Thorsten Holz} {and}
  \bibinfo{person}{Stefan Savage}} (Eds.). \bibinfo{pages}{857--874}.
\newblock
\urldef\tempurl%
\url{https://www.usenix.org/conference/usenixsecurity16/technical-sessions/presentation/costan}
\showURL{%
\tempurl}


\bibitem[de~Miguel~Beriain et~al\mbox{.}(2022)]%
        {eureg}
\bibfield{author}{\bibinfo{person}{Iñigo de Miguel~Beriain} {et~al\mbox{.}}}
  \bibinfo{year}{2022}\natexlab{}.
\newblock \bibinfo{booktitle}{\emph{Auditing the quality of datasets used in
  algorithmic decision-making systems}}.
\newblock \bibinfo{type}{{T}echnical {R}eport}. \bibinfo{institution}{European
  Parliamentary Research Service}.
\newblock
\newblock
\shownote{PE~729.541}.


\bibitem[Duddu et~al\mbox{.}(2024)]%
        {duddu2023attesting}
\bibfield{author}{\bibinfo{person}{Vasisht Duddu} {et~al\mbox{.}}}
  \bibinfo{year}{2024}\natexlab{}.
\newblock \showarticletitle{Attesting Distributional Properties of Training
  Data for Machine Learning}. In \bibinfo{booktitle}{\emph{ESORICS}}.
  \bibinfo{publisher}{Springer}, \bibinfo{address}{Cham}.
\newblock


\bibitem[Eisenhofer et~al\mbox{.}(2022)]%
        {eisenhofer2022verifiable}
\bibfield{author}{\bibinfo{person}{Thorsten Eisenhofer} {et~al\mbox{.}}}
  \bibinfo{year}{2022}\natexlab{}.
\newblock \showarticletitle{Verifiable and provably secure machine unlearning}.
\newblock \bibinfo{journal}{\emph{arXiv preprint arXiv:2210.09126}}
  (\bibinfo{year}{2022}).
\newblock


\bibitem[Fang et~al\mbox{.}(2023)]%
        {PoLbroken}
\bibfield{author}{\bibinfo{person}{C. Fang} {et~al\mbox{.}}}
  \bibinfo{year}{2023}\natexlab{}.
\newblock \showarticletitle{Proof-of-Learning is Currently More Broken Than You
  Think}. In \bibinfo{booktitle}{\emph{EuroS\&P}}. \bibinfo{publisher}{IEEE},
  \bibinfo{address}{Los Alamitos, CA, USA}, \bibinfo{pages}{797--816}.
\newblock
\urldef\tempurl%
\url{https://doi.org/10.1109/EuroSP57164.2023.00052}
\showDOI{\tempurl}


\bibitem[Feng et~al\mbox{.}(2021)]%
        {zenCompiler}
\bibfield{author}{\bibinfo{person}{Boyuan Feng} {et~al\mbox{.}}}
  \bibinfo{year}{2021}\natexlab{}.
\newblock \bibinfo{title}{{ZEN}: An Optimizing Compiler for Verifiable,
  Zero-Knowledge Neural Network Inferences}.
\newblock \bibinfo{howpublished}{Cryptology ePrint Archive, Paper 2021/087}.
\newblock
\urldef\tempurl%
\url{https://eprint.iacr.org/2021/087}
\showURL{%
\tempurl}
\newblock
\shownote{\url{https://eprint.iacr.org/2021/087}}.


\bibitem[Garg et~al\mbox{.}(2023)]%
        {zkpTraining}
\bibfield{author}{\bibinfo{person}{Sanjam Garg} {et~al\mbox{.}}}
  \bibinfo{year}{2023}\natexlab{}.
\newblock \showarticletitle{Experimenting with Zero-Knowledge Proofs of
  Training}. In \bibinfo{booktitle}{\emph{CCS}} (Copenhagen, Denmark).
  \bibinfo{publisher}{Association for Computing Machinery},
  \bibinfo{address}{New York, NY, USA}, \bibinfo{pages}{1880–1894}.
\newblock
\showISBNx{9798400700507}
\urldef\tempurl%
\url{https://doi.org/10.1145/3576915.3623202}
\showDOI{\tempurl}


\bibitem[Gebru et~al\mbox{.}(2021)]%
        {gebru2021datasheets}
\bibfield{author}{\bibinfo{person}{Timnit Gebru} {et~al\mbox{.}}}
  \bibinfo{year}{2021}\natexlab{}.
\newblock \showarticletitle{Datasheets for datasets}.
\newblock \bibinfo{journal}{\emph{Commun. ACM}} \bibinfo{volume}{64},
  \bibinfo{number}{12} (\bibinfo{year}{2021}), \bibinfo{pages}{86--92}.
\newblock


\bibitem[Ghodsi et~al\mbox{.}(2017)]%
        {safetyNets}
\bibfield{author}{\bibinfo{person}{Zahra Ghodsi} {et~al\mbox{.}}}
  \bibinfo{year}{2017}\natexlab{}.
\newblock \showarticletitle{{SafetyNets}: Verifiable Execution of Deep Neural
  Networks on an Untrusted Cloud}. In \bibinfo{booktitle}{\emph{NeurIPS}} (Long
  Beach, California, USA). \bibinfo{publisher}{Curran Associates Inc.},
  \bibinfo{address}{Red Hook, NY, USA}, \bibinfo{pages}{4675–4684}.
\newblock
\showISBNx{9781510860964}


\bibitem[Google(2024)]%
        {modelcardsGoogleCloud}
\bibfield{author}{\bibinfo{person}{Google}.} \bibinfo{year}{2024}\natexlab{}.
\newblock \showarticletitle{{G}oogle {C}loud {M}odel {C}ards ---
  modelcards.withgoogle.com}.
\newblock \bibinfo{journal}{\emph{\url{https://modelcards.withgoogle.com/}}}
  (\bibinfo{year}{2024}).
\newblock
\newblock
\shownote{[Accessed 11-04-2024]}.


\bibitem[Guidotti et~al\mbox{.}(2018)]%
        {explSurvey}
\bibfield{author}{\bibinfo{person}{Riccardo Guidotti} {et~al\mbox{.}}}
  \bibinfo{year}{2018}\natexlab{}.
\newblock \showarticletitle{A Survey of Methods for Explaining Black Box
  Models}.
\newblock \bibinfo{journal}{\emph{ACM Comput. Surv.}} \bibinfo{volume}{51},
  \bibinfo{number}{5}, Article \bibinfo{articleno}{93} (\bibinfo{date}{aug}
  \bibinfo{year}{2018}), \bibinfo{numpages}{42}~pages.
\newblock
\showISSN{0360-0300}
\urldef\tempurl%
\url{https://doi.org/10.1145/3236009}
\showDOI{\tempurl}


\bibitem[Hardt et~al\mbox{.}(2016)]%
        {hardt}
\bibfield{author}{\bibinfo{person}{Moritz Hardt} {et~al\mbox{.}}}
  \bibinfo{year}{2016}\natexlab{}.
\newblock \showarticletitle{Equality of opportunity in supervised learning}. In
  \bibinfo{booktitle}{\emph{NeurIPS}} (Barcelona, Spain).
  \bibinfo{publisher}{Curran Associates Inc.}, \bibinfo{address}{Red Hook, NY,
  USA}, \bibinfo{pages}{3323–3331}.
\newblock
\showISBNx{9781510838819}


\bibitem[{High-Level Expert Group on AI}(2019)]%
        {ec2019ethics}
\bibfield{author}{\bibinfo{person}{{High-Level Expert Group on AI}}.}
  \bibinfo{year}{2019}\natexlab{}.
\newblock \bibinfo{booktitle}{\emph{Ethics guidelines for trustworthy {AI}}}.
\newblock \bibinfo{type}{Report}. \bibinfo{institution}{European Commission},
  \bibinfo{address}{Brussels}.
\newblock
\urldef\tempurl%
\url{https://ec.europa.eu/digital-single-market/en/news/ethics-guidelines-trustworthy-ai}
\showURL{%
\tempurl}


\bibitem[House(2020)]%
        {whitehouse}
\bibfield{author}{\bibinfo{person}{White House}.}
  \bibinfo{year}{2020}\natexlab{}.
\newblock \showarticletitle{Guidance for Regulation of Artificial Intelligence
  Applications}.
\newblock \bibinfo{journal}{\emph{Memorandum For The Heads Of Executive
  Departments And Agencies}} (\bibinfo{year}{2020}).
\newblock
\urldef\tempurl%
\url{https://www.whitehouse.gov/wp-content/uploads/2020/11/M-21-06.pdf}
\showURL{%
\tempurl}


\bibitem[Huang et~al\mbox{.}(2022)]%
        {zkMLaaS}
\bibfield{author}{\bibinfo{person}{Chenyu Huang} {et~al\mbox{.}}}
  \bibinfo{year}{2022}\natexlab{}.
\newblock \showarticletitle{zkMLaaS: a Verifiable Scheme for Machine Learning
  as a Service}. In \bibinfo{booktitle}{\emph{GLOBECOM}}.
  \bibinfo{publisher}{IEEE}, \bibinfo{address}{Los Alamitos, CA, USA},
  \bibinfo{pages}{5475--5480}.
\newblock
\urldef\tempurl%
\url{https://doi.org/10.1109/GLOBECOM48099.2022.10000784}
\showDOI{\tempurl}


\bibitem[HuggingFace(2024a)]%
        {huggingfaceDatasetCards}
\bibfield{author}{\bibinfo{person}{HuggingFace}.}
  \bibinfo{year}{2024}\natexlab{a}.
\newblock \showarticletitle{{D}ataset {C}ards --- huggingface.co}.
\newblock
  \bibinfo{journal}{\emph{\url{https://huggingface.co/docs/hub/en/datasets-cards}}}
  (\bibinfo{year}{2024}).
\newblock
\newblock
\shownote{[Accessed 11-05-2024]}.


\bibitem[HuggingFace(2024b)]%
        {huggingfaceModelCards}
\bibfield{author}{\bibinfo{person}{HuggingFace}.}
  \bibinfo{year}{2024}\natexlab{b}.
\newblock \showarticletitle{{M}odel {C}ards --- huggingface.co}.
\newblock
  \bibinfo{journal}{\emph{\url{https://huggingface.co/docs/hub/en/model-cards}}}
  (\bibinfo{year}{2024}).
\newblock
\newblock
\shownote{[Accessed 11-04-2024]}.


\bibitem[Intel(2014)]%
        {sgx}
\bibfield{author}{\bibinfo{person}{Intel}.} \bibinfo{year}{2014}\natexlab{}.
\newblock \bibinfo{booktitle}{\emph{{Intel Software Guard Extensions
  Programming Reference}}}.
\newblock \bibinfo{type}{Documentation}. \bibinfo{institution}{Intel}.
\newblock
\urldef\tempurl%
\url{https://www.intel.com/content/dam/develop/external/us/en/documents/329298-002-629101.pdf}
\showURL{%
\tempurl}


\bibitem[Intel(2019)]%
        {dcap}
\bibfield{author}{\bibinfo{person}{Intel}.} \bibinfo{year}{2019}\natexlab{}.
\newblock \bibinfo{booktitle}{\emph{{Intel SGX Data Center Attestation
  Primitives (Intel SGX DCAP)}}}.
\newblock \bibinfo{type}{Product Brief}. \bibinfo{institution}{Intel}.
\newblock
\urldef\tempurl%
\url{https://www.intel.com/content/dam/develop/public/us/en/documents/intel-sgx-dcap-ecdsa-orientation.pdf}
\showURL{%
\tempurl}


\bibitem[Intel(2024)]%
        {amx}
\bibfield{author}{\bibinfo{person}{Intel}.} \bibinfo{year}{2024}\natexlab{}.
\newblock \bibinfo{booktitle}{\emph{Accelerate {AI} Workloads with {Intel
  Advanced Matrix Extensions (Intel AMX)}}}.
\newblock \bibinfo{type}{Solution Brief}. \bibinfo{institution}{Intel}.
\newblock


\bibitem[Jagielski et~al\mbox{.}(2020)]%
        {jagielski2020auditing}
\bibfield{author}{\bibinfo{person}{Matthew Jagielski} {et~al\mbox{.}}}
  \bibinfo{year}{2020}\natexlab{}.
\newblock \showarticletitle{Auditing differentially private machine learning:
  How private is private {SGD}?}
\newblock \bibinfo{journal}{\emph{NeurIPS}}  \bibinfo{volume}{33}
  (\bibinfo{year}{2020}), \bibinfo{pages}{22205--22216}.
\newblock


\bibitem[Jia et~al\mbox{.}(2021)]%
        {jia2021proof}
\bibfield{author}{\bibinfo{person}{Hengrui Jia} {et~al\mbox{.}}}
  \bibinfo{year}{2021}\natexlab{}.
\newblock \showarticletitle{Proof-of-Learning: Definitions and Practice}. In
  \bibinfo{booktitle}{\emph{SP}}. \bibinfo{publisher}{IEEE},
  \bibinfo{address}{Los Alamitos, CA, USA}, \bibinfo{pages}{1039--1056}.
\newblock
\urldef\tempurl%
\url{https://doi.org/10.1109/SP40001.2021.00106}
\showDOI{\tempurl}


\bibitem[Jin et~al\mbox{.}(2022)]%
        {jin2022input}
\bibfield{author}{\bibinfo{person}{Jiayin Jin} {et~al\mbox{.}}}
  \bibinfo{year}{2022}\natexlab{}.
\newblock \showarticletitle{Input-agnostic certified group fairness via
  gaussian parameter smoothing}. In \bibinfo{booktitle}{\emph{ICML}}.
  \bibinfo{pages}{10340--10361}.
\newblock


\bibitem[Juarez et~al\mbox{.}(2022)]%
        {juarez2022black}
\bibfield{author}{\bibinfo{person}{Marc Juarez} {et~al\mbox{.}}}
  \bibinfo{year}{2022}\natexlab{}.
\newblock \showarticletitle{Black-Box Audits for Group Distribution Shifts}.
\newblock \bibinfo{journal}{\emph{arXiv preprint arXiv:2209.03620}}
  (\bibinfo{year}{2022}).
\newblock


\bibitem[Kaur et~al\mbox{.}(2022)]%
        {trustAISurvey}
\bibfield{author}{\bibinfo{person}{Davinder Kaur} {et~al\mbox{.}}}
  \bibinfo{year}{2022}\natexlab{}.
\newblock \showarticletitle{Trustworthy Artificial Intelligence: A Review}.
\newblock \bibinfo{journal}{\emph{Comput. Surveys}} \bibinfo{volume}{55},
  \bibinfo{number}{2}, Article \bibinfo{articleno}{39} (\bibinfo{year}{2022}),
  \bibinfo{numpages}{38}~pages.
\newblock
\showISSN{0360-0300}
\urldef\tempurl%
\url{https://doi.org/10.1145/3491209}
\showDOI{\tempurl}


\bibitem[Kazim et~al\mbox{.}(2021)]%
        {ico}
\bibfield{author}{\bibinfo{person}{Emre Kazim} {et~al\mbox{.}}}
  \bibinfo{year}{2021}\natexlab{}.
\newblock \showarticletitle{{AI} auditing and impact assessment: according to
  the {UK} information commissioner’s office}.
\newblock \bibinfo{journal}{\emph{AI and Ethics}} (\bibinfo{date}{Feb}
  \bibinfo{year}{2021}).
\newblock
\showISSN{2730-5953, 2730-5961}
\urldef\tempurl%
\url{https://doi.org/10.1007/s43681-021-00039-2}
\showDOI{\tempurl}


\bibitem[Khedr and Shoukry(2023)]%
        {khedr2023certifair}
\bibfield{author}{\bibinfo{person}{Haitham Khedr} {and} \bibinfo{person}{Yasser
  Shoukry}.} \bibinfo{year}{2023}\natexlab{}.
\newblock \showarticletitle{CertiFair: a framework for certified global
  fairness of neural networks}. In \bibinfo{booktitle}{\emph{AAAI}}.
  \bibinfo{publisher}{AAAI Press}, Article \bibinfo{articleno}{925},
  \bibinfo{numpages}{9}~pages.
\newblock
\showISBNx{978-1-57735-880-0}
\urldef\tempurl%
\url{https://doi.org/10.1609/aaai.v37i7.25994}
\showDOI{\tempurl}


\bibitem[Kilbertus et~al\mbox{.}(2018)]%
        {fairaudit1}
\bibfield{author}{\bibinfo{person}{Niki Kilbertus}, \bibinfo{person}{Adria
  Gascon}, \bibinfo{person}{Matt Kusner}, \bibinfo{person}{Michael Veale},
  \bibinfo{person}{Krishna Gummadi}, {and} \bibinfo{person}{Adrian Weller}.}
  \bibinfo{year}{2018}\natexlab{}.
\newblock \showarticletitle{Blind Justice: Fairness with Encrypted Sensitive
  Attributes}. In \bibinfo{booktitle}{\emph{ICML}}, Vol.~\bibinfo{volume}{80}.
  \bibinfo{publisher}{PMLR}, \bibinfo{pages}{2630--2639}.
\newblock
\urldef\tempurl%
\url{https://proceedings.mlr.press/v80/kilbertus18a.html}
\showURL{%
\tempurl}


\bibitem[Kohavi(1996)]%
        {census}
\bibfield{author}{\bibinfo{person}{Ron Kohavi}.}
  \bibinfo{year}{1996}\natexlab{}.
\newblock \bibinfo{title}{{Census Income}}.
\newblock \bibinfo{howpublished}{UCI Machine Learning Repository}.
\newblock
\newblock
\shownote{{DOI}: https://doi.org/10.24432/C5GP7S}.


\bibitem[Kong et~al\mbox{.}(2022)]%
        {miaRepudiation}
\bibfield{author}{\bibinfo{person}{Zhifeng Kong} {et~al\mbox{.}}}
  \bibinfo{year}{2022}\natexlab{}.
\newblock \showarticletitle{Forgeability and Membership Inference Attacks}. In
  \bibinfo{booktitle}{\emph{AISec}} (Los Angeles, CA, USA).
  \bibinfo{publisher}{Association for Computing Machinery},
  \bibinfo{address}{New York, NY, USA}, \bibinfo{pages}{25–31}.
\newblock
\showISBNx{9781450398800}
\urldef\tempurl%
\url{https://doi.org/10.1145/3560830.3563731}
\showDOI{\tempurl}


\bibitem[Kostiainen et~al\mbox{.}(2010)]%
        {tee}
\bibfield{author}{\bibinfo{person}{Kari Kostiainen} {et~al\mbox{.}}}
  \bibinfo{year}{2010}\natexlab{}.
\newblock \showarticletitle{Key Attestation from Trusted Execution
  Environments}. In \bibinfo{booktitle}{\emph{TRUST}}.
\newblock
\urldef\tempurl%
\url{https://doi.org/10.1007/978-3-642-13869-0_3}
\showDOI{\tempurl}


\bibitem[Kuvaiskii et~al\mbox{.}(2022)]%
        {kuvaiskii2022computation}
\bibfield{author}{\bibinfo{person}{Dmitrii Kuvaiskii} {et~al\mbox{.}}}
  \bibinfo{year}{2022}\natexlab{}.
\newblock \showarticletitle{Computation offloading to hardware accelerators in
  {Intel SGX} and {Gramine} library {OS}}.
\newblock \bibinfo{journal}{\emph{arXiv preprint arXiv:2203.01813}}
  (\bibinfo{year}{2022}).
\newblock


\bibitem[Law(2018)]%
        {dpia}
\bibfield{author}{\bibinfo{person}{European~Union Law}.}
  \bibinfo{year}{2018}\natexlab{}.
\newblock \showarticletitle{Art. 35 {GDPR} Data protection impact assessment}.
\newblock \bibinfo{journal}{\emph{General Data Protection Regulation (GDPR)}}
  (\bibinfo{year}{2018}).
\newblock
\urldef\tempurl%
\url{https://gdpr-info.eu/art-35-gdpr/}
\showURL{%
\tempurl}


\bibitem[Li and Xothers(2023)]%
        {li2023sok}
\bibfield{author}{\bibinfo{person}{Linyi Li} {and} \bibinfo{person}{Xothers}.}
  \bibinfo{year}{2023}\natexlab{}.
\newblock \showarticletitle{{SoK}: Certified Robustness for Deep Neural
  Networks}. In \bibinfo{booktitle}{\emph{SP}}. \bibinfo{publisher}{IEEE
  Computer Society}, \bibinfo{address}{Los Alamitos, CA, USA},
  \bibinfo{pages}{1289--1310}.
\newblock
\urldef\tempurl%
\url{https://doi.org/10.1109/SP46215.2023.10179303}
\showDOI{\tempurl}


\bibitem[Liu et~al\mbox{.}(2021)]%
        {zkCNN}
\bibfield{author}{\bibinfo{person}{Tianyi Liu} {et~al\mbox{.}}}
  \bibinfo{year}{2021}\natexlab{}.
\newblock \showarticletitle{zkCNN: Zero Knowledge Proofs for Convolutional
  Neural Network Predictions and Accuracy}. In \bibinfo{booktitle}{\emph{CCS}}
  (Virtual Event, Republic of Korea). \bibinfo{publisher}{Association for
  Computing Machinery}, \bibinfo{address}{New York, NY, USA},
  \bibinfo{pages}{2968–2985}.
\newblock
\showISBNx{9781450384544}
\urldef\tempurl%
\url{https://doi.org/10.1145/3460120.3485379}
\showDOI{\tempurl}


\bibitem[Lycklama et~al\mbox{.}(2024)]%
        {lycklama2024holding}
\bibfield{author}{\bibinfo{person}{Hidde Lycklama} {et~al\mbox{.}}}
  \bibinfo{year}{2024}\natexlab{}.
\newblock \showarticletitle{Holding Secrets Accountable: Auditing
  Privacy-Preserving Machine Learning}. In \bibinfo{booktitle}{\emph{USENIX
  Sec}}. \bibinfo{publisher}{USENIX Association}, \bibinfo{address}{Baltimore,
  MD}, \bibinfo{pages}{991{\textendash}1008}.
\newblock
\urldef\tempurl%
\url{https://www.usenix.org/conference/usenixsecurity24/presentation/lycklama}
\showURL{%
\tempurl}


\bibitem[Maas et~al\mbox{.}(2011)]%
        {imdb}
\bibfield{author}{\bibinfo{person}{Andrew~L. Maas} {et~al\mbox{.}}}
  \bibinfo{year}{2011}\natexlab{}.
\newblock \showarticletitle{Learning Word Vectors for Sentiment Analysis}. In
  \bibinfo{booktitle}{\emph{ACL}}. \bibinfo{publisher}{Association for
  Computational Linguistics}, \bibinfo{address}{Portland, Oregon, USA},
  \bibinfo{pages}{142--150}.
\newblock
\urldef\tempurl%
\url{http://www.aclweb.org/anthology/P11-1015}
\showURL{%
\tempurl}


\bibitem[Miao et~al\mbox{.}(2021)]%
        {miao2021audio}
\bibfield{author}{\bibinfo{person}{Yuantian Miao} {et~al\mbox{.}}}
  \bibinfo{year}{2021}\natexlab{}.
\newblock \showarticletitle{The audio auditor: user-level membership inference
  in internet of things voice services}.
\newblock \bibinfo{journal}{\emph{PoPETS}}  \bibinfo{volume}{2021}
  (\bibinfo{year}{2021}), \bibinfo{pages}{209--228}.
\newblock


\bibitem[Mitchell et~al\mbox{.}(2019)]%
        {mitchell2019model}
\bibfield{author}{\bibinfo{person}{Margaret Mitchell} {et~al\mbox{.}}}
  \bibinfo{year}{2019}\natexlab{}.
\newblock \showarticletitle{Model Cards for Model Reporting}. In
  \bibinfo{booktitle}{\emph{FaccT}} (Atlanta, GA, USA).
  \bibinfo{publisher}{Association for Computing Machinery},
  \bibinfo{address}{New York, NY, USA}, \bibinfo{pages}{220–229}.
\newblock
\showISBNx{9781450361255}
\urldef\tempurl%
\url{https://doi.org/10.1145/3287560.3287596}
\showDOI{\tempurl}


\bibitem[Mithril-Security(2023)]%
        {mithrilsecurityPoisonGPTPoison}
\bibfield{author}{\bibinfo{person}{Mithril-Security}.}
  \bibinfo{year}{2023}\natexlab{}.
\newblock \bibinfo{title}{{P}oison{G}{P}{T}: {H}ow to poison {L}{L}{M} supply
  chain on {H}ugging {F}ace --- blog.mithrilsecurity.io}.
\newblock
  \bibinfo{howpublished}{https://blog.mithrilsecurity.io/poisongpt-how-we-hid-a-lobotomized-llm-on-hugging-face-to-spread-fake-news/}.
\newblock
\newblock
\shownote{[Accessed 10-05-2024]}.


\bibitem[Mo et~al\mbox{.}(2020)]%
        {mo2020darknetz}
\bibfield{author}{\bibinfo{person}{Fan Mo} {et~al\mbox{.}}}
  \bibinfo{year}{2020}\natexlab{}.
\newblock \showarticletitle{DarkneTZ: towards model privacy at the edge using
  trusted execution environments}. In \bibinfo{booktitle}{\emph{MobiSys}}
  (Toronto, Ontario, Canada). \bibinfo{publisher}{Association for Computing
  Machinery}, \bibinfo{address}{New York, NY, USA}, \bibinfo{pages}{161–174}.
\newblock
\showISBNx{9781450379540}
\urldef\tempurl%
\url{https://doi.org/10.1145/3386901.3388946}
\showDOI{\tempurl}


\bibitem[Mo et~al\mbox{.}(2021)]%
        {mo2021ppfl}
\bibfield{author}{\bibinfo{person}{Fan Mo} {et~al\mbox{.}}}
  \bibinfo{year}{2021}\natexlab{}.
\newblock \showarticletitle{PPFL: privacy-preserving federated learning with
  trusted execution environments}. In \bibinfo{booktitle}{\emph{MobiSys}}
  (Virtual Event, Wisconsin). \bibinfo{publisher}{Association for Computing
  Machinery}, \bibinfo{address}{New York, NY, USA}, \bibinfo{pages}{94–108}.
\newblock
\showISBNx{9781450384438}
\urldef\tempurl%
\url{https://doi.org/10.1145/3458864.3466628}
\showDOI{\tempurl}


\bibitem[Mo et~al\mbox{.}(2022)]%
        {mo2022sok}
\bibfield{author}{\bibinfo{person}{Fan Mo} {et~al\mbox{.}}}
  \bibinfo{year}{2022}\natexlab{}.
\newblock \showarticletitle{{SoK}: machine learning with confidential
  computing}.
\newblock \bibinfo{journal}{\emph{arXiv preprint arXiv:2208.10134}}
  (\bibinfo{year}{2022}).
\newblock


\bibitem[Nasr et~al\mbox{.}(2023)]%
        {nasr2023tight}
\bibfield{author}{\bibinfo{person}{Milad Nasr} {et~al\mbox{.}}}
  \bibinfo{year}{2023}\natexlab{}.
\newblock \showarticletitle{Tight auditing of differentially private machine
  learning}. In \bibinfo{booktitle}{\emph{USENIX Sec}}.
  \bibinfo{pages}{1631--1648}.
\newblock


\bibitem[Ng et~al\mbox{.}(2021)]%
        {goten}
\bibfield{author}{\bibinfo{person}{Lucien K.~L. Ng} {et~al\mbox{.}}}
  \bibinfo{year}{2021}\natexlab{}.
\newblock \showarticletitle{Goten: GPU-Outsourcing Trusted Execution of Neural
  Network Training}.
\newblock \bibinfo{journal}{\emph{AAAI}} \bibinfo{volume}{35},
  \bibinfo{number}{17} (\bibinfo{date}{May} \bibinfo{year}{2021}),
  \bibinfo{pages}{14876--14883}.
\newblock
\urldef\tempurl%
\url{https://doi.org/10.1609/aaai.v35i17.17746}
\showDOI{\tempurl}


\bibitem[Ng and Chow(2023)]%
        {SoKCrypto}
\bibfield{author}{\bibinfo{person}{Lucien K.~L. Ng} {and}
  \bibinfo{person}{Sherman S.~M. Chow}.} \bibinfo{year}{2023}\natexlab{}.
\newblock \showarticletitle{SoK: Cryptographic Neural-Network Computation}. In
  \bibinfo{booktitle}{\emph{2023 IEEE Symposium on Security and Privacy (SP)}}.
  \bibinfo{pages}{497--514}.
\newblock
\urldef\tempurl%
\url{https://doi.org/10.1109/SP46215.2023.10179483}
\showDOI{\tempurl}


\bibitem[Nvidia(2024)]%
        {h100}
\bibfield{author}{\bibinfo{person}{Nvidia}.} \bibinfo{year}{2024}\natexlab{}.
\newblock \bibinfo{booktitle}{\emph{{NVIDIA} {H100} Tensor Core {GPU}}}.
\newblock \bibinfo{type}{Datasheet}. \bibinfo{institution}{Nvidia}.
\newblock


\bibitem[Park et~al\mbox{.}(2022)]%
        {fairaudit2}
\bibfield{author}{\bibinfo{person}{Saerom Park} {et~al\mbox{.}}}
  \bibinfo{year}{2022}\natexlab{}.
\newblock \showarticletitle{Fairness Audit of Machine Learning Models with
  Confidential Computing}. In \bibinfo{booktitle}{\emph{WWW}} (Virtual Event,
  Lyon, France). \bibinfo{publisher}{Association for Computing Machinery},
  \bibinfo{address}{New York, NY, USA}, \bibinfo{pages}{3488–3499}.
\newblock
\showISBNx{9781450390965}
\urldef\tempurl%
\url{https://doi.org/10.1145/3485447.3512244}
\showDOI{\tempurl}


\bibitem[Paszke et~al\mbox{.}(2019)]%
        {paszke2019pytorch}
\bibfield{author}{\bibinfo{person}{Adam Paszke} {et~al\mbox{.}}}
  \bibinfo{year}{2019}\natexlab{}.
\newblock \showarticletitle{Pytorch: An imperative style, high-performance deep
  learning library}.
\newblock \bibinfo{journal}{\emph{NeurIPS}}  \bibinfo{volume}{32}
  (\bibinfo{year}{2019}).
\newblock


\bibitem[Pushkarna et~al\mbox{.}(2022)]%
        {pushkarna2022data}
\bibfield{author}{\bibinfo{person}{Mahima Pushkarna} {et~al\mbox{.}}}
  \bibinfo{year}{2022}\natexlab{}.
\newblock \showarticletitle{Data Cards: Purposeful and Transparent Dataset
  Documentation for Responsible AI}. In \bibinfo{booktitle}{\emph{FaccT}}
  (Seoul, Republic of Korea). \bibinfo{publisher}{Association for Computing
  Machinery}, \bibinfo{address}{New York, NY, USA},
  \bibinfo{pages}{1776–1826}.
\newblock
\showISBNx{9781450393522}
\urldef\tempurl%
\url{https://doi.org/10.1145/3531146.3533231}
\showDOI{\tempurl}


\bibitem[Rezaei and Liu(2021)]%
        {rezaei2021difficulty}
\bibfield{author}{\bibinfo{person}{Shahbaz Rezaei} {and} \bibinfo{person}{Xin
  Liu}.} \bibinfo{year}{2021}\natexlab{}.
\newblock \showarticletitle{On the Difficulty of Membership Inference Attacks}.
  In \bibinfo{booktitle}{\emph{CVPR}}. \bibinfo{publisher}{IEEE},
  \bibinfo{address}{Los Alamitos, CA, USA}, \bibinfo{pages}{7892--7900}.
\newblock


\bibitem[Sadeghi and St\"{u}ble(2004)]%
        {propatt}
\bibfield{author}{\bibinfo{person}{Ahmad-Reza Sadeghi} {and}
  \bibinfo{person}{Christian St\"{u}ble}.} \bibinfo{year}{2004}\natexlab{}.
\newblock \showarticletitle{Property-based attestation for computing platforms:
  caring about properties, not mechanisms}. In \bibinfo{booktitle}{\emph{NSPW}}
  (Nova Scotia, Canada). \bibinfo{publisher}{Association for Computing
  Machinery}, \bibinfo{address}{New York, NY, USA}, \bibinfo{pages}{67–77}.
\newblock
\showISBNx{1595930760}
\urldef\tempurl%
\url{https://doi.org/10.1145/1065907.1066038}
\showDOI{\tempurl}


\bibitem[Sardar and Fetzer(2023)]%
        {confidential-computing}
\bibfield{author}{\bibinfo{person}{Muhammad~Usama Sardar} {and}
  \bibinfo{person}{Christof Fetzer}.} \bibinfo{year}{2023}\natexlab{}.
\newblock \showarticletitle{Confidential computing and related technologies: a
  critical review}.
\newblock \bibinfo{journal}{\emph{Cybersecur.}} \bibinfo{volume}{6},
  \bibinfo{number}{1} (\bibinfo{year}{2023}), \bibinfo{pages}{10}.
\newblock
\urldef\tempurl%
\url{https://doi.org/10.1186/S42400-023-00144-1}
\showDOI{\tempurl}


\bibitem[Segal et~al\mbox{.}(2021)]%
        {fairaudit3}
\bibfield{author}{\bibinfo{person}{Shahar Segal} {et~al\mbox{.}}}
  \bibinfo{year}{2021}\natexlab{}.
\newblock \showarticletitle{Fairness in the Eyes of the Data: Certifying
  Machine-Learning Models}. In \bibinfo{booktitle}{\emph{AIES}} (Virtual Event,
  USA). \bibinfo{publisher}{Association for Computing Machinery},
  \bibinfo{address}{New York, NY, USA}, \bibinfo{pages}{926–935}.
\newblock
\showISBNx{9781450384735}
\urldef\tempurl%
\url{https://doi.org/10.1145/3461702.3462554}
\showDOI{\tempurl}


\bibitem[Shamsabadi et~al\mbox{.}(2023)]%
        {shamsabadi2023confidentialprofitt}
\bibfield{author}{\bibinfo{person}{Ali~Shahin Shamsabadi} {et~al\mbox{.}}}
  \bibinfo{year}{2023}\natexlab{}.
\newblock \showarticletitle{Confidential-{PROFITT}: {Confidential} {PRO}of of
  {FaIr} {Training} of {Trees}}. In \bibinfo{booktitle}{\emph{ICLR}}.
\newblock
\urldef\tempurl%
\url{https://openreview.net/forum?id=iIfDQVyuFD}
\showURL{%
\tempurl}


\bibitem[Shamsabadi et~al\mbox{.}(2024)]%
        {confidentialdpproof}
\bibfield{author}{\bibinfo{person}{Ali~Shahin Shamsabadi} {et~al\mbox{.}}}
  \bibinfo{year}{2024}\natexlab{}.
\newblock \showarticletitle{Confidential-{DP}proof: Confidential Proof of
  Differentially Private Training}. In \bibinfo{booktitle}{\emph{ICLR}}.
\newblock
\urldef\tempurl%
\url{https://openreview.net/forum?id=PQY2v6VtGe}
\showURL{%
\tempurl}


\bibitem[Simonyan(2014)]%
        {vgg}
\bibfield{author}{\bibinfo{person}{Karen Simonyan}.}
  \bibinfo{year}{2014}\natexlab{}.
\newblock \showarticletitle{Very deep convolutional networks for large-scale
  image recognition}.
\newblock \bibinfo{journal}{\emph{arXiv preprint arXiv:1409.1556}}
  (\bibinfo{year}{2014}).
\newblock


\bibitem[Song and Shmatikov(2019)]%
        {auditingMIA}
\bibfield{author}{\bibinfo{person}{Congzheng Song} {and}
  \bibinfo{person}{Vitaly Shmatikov}.} \bibinfo{year}{2019}\natexlab{}.
\newblock \showarticletitle{Auditing Data Provenance in Text-Generation
  Models}. In \bibinfo{booktitle}{\emph{KDD}} (Anchorage, AK, USA).
  \bibinfo{publisher}{Association for Computing Machinery},
  \bibinfo{address}{New York, NY, USA}, \bibinfo{pages}{196--206}.
\newblock
\showISBNx{9781450362016}
\urldef\tempurl%
\url{https://doi.org/10.1145/3292500.3330885}
\showDOI{\tempurl}


\bibitem[Srivastava et~al\mbox{.}(2024)]%
        {srivastava2024optimistic}
\bibfield{author}{\bibinfo{person}{Megha Srivastava} {et~al\mbox{.}}}
  \bibinfo{year}{2024}\natexlab{}.
\newblock \showarticletitle{Optimistic Verifiable Training by Controlling
  Hardware Nondeterminism}. In \bibinfo{booktitle}{\emph{The Thirty-eighth
  Annual Conference on Neural Information Processing Systems}}.
\newblock
\urldef\tempurl%
\url{https://openreview.net/forum?id=bf0MdFlz1i}
\showURL{%
\tempurl}


\bibitem[Steinke et~al\mbox{.}(2024)]%
        {steinke2024privacy}
\bibfield{author}{\bibinfo{person}{Thomas Steinke} {et~al\mbox{.}}}
  \bibinfo{year}{2024}\natexlab{}.
\newblock \showarticletitle{Privacy auditing with one (1) training run}.
\newblock \bibinfo{journal}{\emph{NeurIPS}}  \bibinfo{volume}{36}
  (\bibinfo{year}{2024}).
\newblock


\bibitem[Sun et~al\mbox{.}(2024)]%
        {sun2024zkllm}
\bibfield{author}{\bibinfo{person}{Haochen Sun} {et~al\mbox{.}}}
  \bibinfo{year}{2024}\natexlab{}.
\newblock \showarticletitle{{zkLLM}: Zero Knowledge Proofs for Large Language
  Models}.
\newblock \bibinfo{journal}{\emph{arXiv preprint arXiv:2404.16109}}
  (\bibinfo{year}{2024}).
\newblock


\bibitem[Sun and Zhang(2023)]%
        {zkDL}
\bibfield{author}{\bibinfo{person}{Haochen Sun} {and} \bibinfo{person}{Hongyang
  Zhang}.} \bibinfo{year}{2023}\natexlab{}.
\newblock \bibinfo{title}{{zkDL}: Efficient Zero-Knowledge Proofs of Deep
  Learning Training}.
\newblock \bibinfo{howpublished}{Cryptology ePrint Archive, Paper 2023/1174}.
\newblock
\urldef\tempurl%
\url{https://eprint.iacr.org/2023/1174}
\showURL{%
\tempurl}
\newblock
\shownote{\url{https://eprint.iacr.org/2023/1174}}.


\bibitem[Tabassi et~al\mbox{.}(2019)]%
        {nist}
\bibfield{author}{\bibinfo{person}{Elham Tabassi} {et~al\mbox{.}}}
  \bibinfo{year}{2019}\natexlab{}.
\newblock \bibinfo{booktitle}{\emph{A Taxonomy and Terminology of Adversarial
  Machine Learning}}.
\newblock \bibinfo{type}{{NIST Interagency/Internal Report}}.
  \bibinfo{institution}{NIST}.
\newblock
\urldef\tempurl%
\url{https://nvlpubs.nist.gov/nistpubs/ir/2019/NIST.IR.8269-draft.pdf}
\showURL{%
\tempurl}


\bibitem[Tramer and Boneh(2019)]%
        {tramer2018slalom}
\bibfield{author}{\bibinfo{person}{Florian Tramer} {and} \bibinfo{person}{Dan
  Boneh}.} \bibinfo{year}{2019}\natexlab{}.
\newblock \showarticletitle{Slalom: Fast, Verifiable and Private Execution of
  Neural Networks in Trusted Hardware}. In \bibinfo{booktitle}{\emph{ICLR}}.
\newblock
\urldef\tempurl%
\url{https://openreview.net/forum?id=rJVorjCcKQ}
\showURL{%
\tempurl}


\bibitem[Urban et~al\mbox{.}(2020)]%
        {urban20}
\bibfield{author}{\bibinfo{person}{Caterina Urban} {et~al\mbox{.}}}
  \bibinfo{year}{2020}\natexlab{}.
\newblock \showarticletitle{Perfectly parallel fairness certification of neural
  networks}.
\newblock \bibinfo{journal}{\emph{Proceedings of the ACM on Programming
  Languages (OOPSLA)}}  \bibinfo{volume}{4}, Article \bibinfo{articleno}{185}
  (\bibinfo{date}{Nov} \bibinfo{year}{2020}), \bibinfo{numpages}{30}~pages.
\newblock


\bibitem[Van~Schaik et~al\mbox{.}(2024)]%
        {sgxfail}
\bibfield{author}{\bibinfo{person}{Stephan Van~Schaik} {et~al\mbox{.}}}
  \bibinfo{year}{2024}\natexlab{}.
\newblock \showarticletitle{{SoK}: SGX.Fail: How Stuff Gets eXposed}. In
  \bibinfo{booktitle}{\emph{SP}}. \bibinfo{publisher}{IEEE Computer Society},
  \bibinfo{address}{Los Alamitos, CA, USA}, \bibinfo{pages}{4143--4162}.
\newblock
\urldef\tempurl%
\url{https://doi.org/10.1109/SP54263.2024.00260}
\showDOI{\tempurl}


\bibitem[Weng et~al\mbox{.}(2021)]%
        {Weng2021MystiqueEC}
\bibfield{author}{\bibinfo{person}{Chenkai Weng} {et~al\mbox{.}}}
  \bibinfo{year}{2021}\natexlab{}.
\newblock \showarticletitle{Mystique: Efficient Conversions for Zero-Knowledge
  Proofs with Applications to Machine Learning}.
\newblock \bibinfo{journal}{\emph{IACR Cryptology ePrint Arch.}}
  \bibinfo{volume}{2021} (\bibinfo{year}{2021}), \bibinfo{pages}{730}.
\newblock
\urldef\tempurl%
\url{https://api.semanticscholar.org/CorpusID:235349056}
\showURL{%
\tempurl}


\bibitem[Yadav et~al\mbox{.}(2024)]%
        {yadav2024fairproof}
\bibfield{author}{\bibinfo{person}{Chhavi Yadav} {et~al\mbox{.}}}
  \bibinfo{year}{2024}\natexlab{}.
\newblock \showarticletitle{FairProof: Confidential and Certifiable Fairness
  for Neural Networks}.
\newblock \bibinfo{journal}{\emph{arXiv preprint arXiv:2402.12572}}
  (\bibinfo{year}{2024}).
\newblock


\bibitem[Yu et~al\mbox{.}(2020)]%
        {stt}
\bibfield{author}{\bibinfo{person}{Jiyong Yu} {et~al\mbox{.}}}
  \bibinfo{year}{2020}\natexlab{}.
\newblock \showarticletitle{Speculative Taint Tracking {(STT):} {A}
  Comprehensive Protection for Speculatively Accessed Data}.
\newblock \bibinfo{journal}{\emph{{IEEE} Micro}} \bibinfo{volume}{40},
  \bibinfo{number}{3} (\bibinfo{year}{2020}), \bibinfo{pages}{81--90}.
\newblock
\urldef\tempurl%
\url{https://doi.org/10.1109/MM.2020.2985359}
\showDOI{\tempurl}


\bibitem[Zhang et~al\mbox{.}(2022)]%
        {zhang2022adversarial}
\bibfield{author}{\bibinfo{person}{R. Zhang} {et~al\mbox{.}}}
  \bibinfo{year}{2022}\natexlab{}.
\newblock \showarticletitle{"Adversarial Examples" for Proof-of-Learning}. In
  \bibinfo{booktitle}{\emph{SP}}. \bibinfo{publisher}{IEEE Computer Society},
  \bibinfo{address}{Los Alamitos, CA, USA}, \bibinfo{pages}{1542--1542}.
\newblock
\showISSN{2375-1207}
\urldef\tempurl%
\url{https://doi.org/10.1109/SP46214.2022.00113}
\showDOI{\tempurl}


\bibitem[Zhang et~al\mbox{.}(2023)]%
        {veriTrain}
\bibfield{author}{\bibinfo{person}{Xiaokuan Zhang} {et~al\mbox{.}}}
  \bibinfo{year}{2023}\natexlab{}.
\newblock \showarticletitle{{VeriTrain}: Validating {MLaaS} Training Efforts
  via Anomaly Detection}.
\newblock \bibinfo{journal}{\emph{IEEE TIFS}} (\bibinfo{year}{2023}),
  \bibinfo{pages}{1--17}.
\newblock
\urldef\tempurl%
\url{https://doi.org/10.1109/TDSC.2023.3266427}
\showDOI{\tempurl}


\bibitem[Zhang et~al\mbox{.}(2017)]%
        {utkface}
\bibfield{author}{\bibinfo{person}{Zhifei Zhang} {et~al\mbox{.}}}
  \bibinfo{year}{2017}\natexlab{}.
\newblock \showarticletitle{Age Progression/Regression by Conditional
  Adversarial Autoencoder}. In \bibinfo{booktitle}{\emph{CVPR}}. IEEE.
\newblock


\bibitem[Zhang et~al\mbox{.}(2024)]%
        {noprivacy}
\bibfield{author}{\bibinfo{person}{Z. Zhang} {et~al\mbox{.}}}
  \bibinfo{year}{2024}\natexlab{}.
\newblock \showarticletitle{No Privacy Left Outside: On the (In-)Security of
  TEE-Shielded DNN Partition for On-Device ML}. In
  \bibinfo{booktitle}{\emph{SP}}. \bibinfo{publisher}{IEEE Computer Society},
  \bibinfo{address}{Los Alamitos, CA, USA}, \bibinfo{pages}{55--55}.
\newblock
\showISSN{2375-1207}
\urldef\tempurl%
\url{https://doi.org/10.1109/SP54263.2024.00052}
\showDOI{\tempurl}


\bibitem[Zhu et~al\mbox{.}(2024)]%
        {H100-performance}
\bibfield{author}{\bibinfo{person}{Jianwei Zhu} {et~al\mbox{.}}}
  \bibinfo{year}{2024}\natexlab{}.
\newblock \showarticletitle{Confidential Computing on NVIDIA Hopper GPUs: A
  Performance Benchmark Study}.
\newblock \bibinfo{journal}{\emph{arXiv preprint arXiv:2409.03992}}
  (\bibinfo{year}{2024}).
\newblock


\end{thebibliography}
